\documentclass[conference]{IEEEtran}
\usepackage{graphicx}
\usepackage{amsthm}
\usepackage{epsfig}
\usepackage{latexsym}
\usepackage{amsfonts}
\usepackage{here}
\usepackage{rawfonts}
\usepackage[latin1]{inputenc}
\usepackage[T1]{fontenc}
\usepackage{calc}
\usepackage{capitalgreekitalic}
\usepackage{url}
\usepackage{enumerate}
\usepackage{color}
\usepackage[tbtags]{amsmath}
\usepackage{amssymb}
\usepackage{upref}
\usepackage{epic,eepic}
\usepackage{times}
\usepackage{dsfont}
\usepackage{comment}
\usepackage{cite}
\usepackage{bbm} 
\usepackage{optidef}
\usepackage{empheq}
\usepackage{lipsum}
\usepackage{setspace}
\usepackage[british]{babel}
\usepackage{enumitem} 

\newtheorem{remark}{\bf Remark}

\usepackage{dsfont}

\newcounter{step}
\newlength{\totlinewidth}
\newenvironment{algorithm}{%
  \rule{\linewidth}{1pt}
  \begin{list}{}%
    {\usecounter{step}%
      \settowidth{\labelwidth}{\textbf{Step 2:}}%
      \setlength{\leftmargin}{\labelwidth}%
      \setlength{\topsep}{-2pt}%
      \addtolength{\leftmargin}{\labelsep}%
      \addtolength{\leftmargin}{2mm}%
      \setlength{\rightmargin}{2mm}%
      \setlength{\totlinewidth}{\linewidth}%
      \addtolength{\totlinewidth}{\leftmargin}%
      \addtolength{\totlinewidth}{\rightmargin}%
      \setlength{\parsep}{0mm}%
      \raggedright}}%
  {\end{list}%
  \rule{\linewidth}{1pt}}
\newcounter{substep}

  {\end{list}}

\newlength{\aligntop}
\newlength{\alignbot}
 

\IEEEoverridecommandlockouts

\usepackage{algorithm}
\usepackage{algorithmic}

\usepackage{subfigure}

\begin{document}
\title{\LARGE Distributed and Distribution-Robust Meta Reinforcement Learning (D$^2$-RMRL) for Data Pre-storing  and Routing in Cube Satellite Networks}

\author{
\IEEEauthorblockN{Ye Hu, Xiaodong Wang, \emph{Fellow, IEEE}, Walid Saad, \emph{Fellow, IEEE} \thanks{{This research was supported by the U.S. National Science Foundation under Grant CNS-1909372.}}} 
\thanks{Y. Hu is with the Department of Electrical Engineering, Columbia University, New York, NY, USA, 24061, and the Wireless@VT, Bradley Department of Electrical and Computer Engineering, Virginia Tech, Blacksburg, VA, USA, 24061, Emails: \protect{yh3453@columbia.edu}.}
\thanks{X. Wang is with the Department of Electrical Engineering, Columbia University, New York, NY, USA, 24061, Email: \protect{xw2008@columbia.edu}.}
\thanks{W. Saad is with the Wireless@VT, Bradley Department of Electrical and Computer Engineering, Virginia Tech, Blacksburg, VA, USA, 24061, Emails: \protect{walids@vt.edu}.}

}



\maketitle
\begin{abstract}
\boldmath
 {In this paper, the problem of data pre-storing and routing in dynamic, resource-constrained cube satellite networks is studied. In such a network, each cube satellite delivers requested data to user clusters under its coverage. 
 A group of ground gateways will route and pre-store certain data to the satellites, such that the ground users can be directly served with the pre-stored data. This pre-storing and routing design problem is formulated as a decentralized Markov decision process (Dec-MDP) in which we seek to find the optimal strategy that maximizes the pre-store hit rate, i.e., the fraction of users being directly served with the pre-stored data. To obtain the optimal strategy, a distributed distribution-robust meta reinforcement learning (D$^2$-RMRL) algorithm is proposed that consists of three key ingredients: value-decomposition for achieving the global optimum in distributed setting with minimum communication overhead, meta learning to obtain the optimal initial to reduce the training time under dynamic conditions, and pre-training to further speed up the meta training procedure.  
Simulation results show that, using the proposed value decomposition and meta training techniques, the satellite networks can achieve a $31.8\%$ improvement of the pre-store hits and a $40.7\%$ improvement of the convergence speed, compared to a baseline reinforcement learning algorithm. Moreover, the use of the proposed pre-training mechanism helps to shorten the meta-learning procedure by up to $43.7\%$.
}
 \end{abstract}


\renewcommand{\thefootnote}{\arabic{footnote}}


%
\IEEEpeerreviewmaketitle

\section{Introduction}
{The low earth orbit (LEO) cube satellite networks can provide an endurable, reliable, and accessible data service to users in wireless disadvantaged areas \cite{maral2020satellite, 7956007}. However, deploying cube satellites for low latency information access is still an important open problem,  because satellites have to serve unforeseeable data needs with their limited available on-board resources. 
In particular, it is challenging for cube satellites to serve unpredictable, diverse data needs from users around the globe using only limited contact chances in the network.

\subsection{Related Works and their Limitations}
Prior works \cite{8906029, 7883913, fraire2021scalability} studied a number of problems related to routing design in resource constrained cube satellite networks. 
The work in \cite{8906029} studies the problem of contact plan design in LEO satellite networks while jointly considering the satellites' on-board energy capacity and their stochastic solar energy infeed. The authors in \cite{7883913} design a contact plan for a resource constrained LEO satellite network to deliver the satellites' data to a fixed ground base station. In \cite{fraire2021scalability}, the problem of scalable battery aware contact plan design in mega LEO satellite constellations is treated using mixed integer linear programing. 
Despite their promising results, these existing on-demand routing solutions \cite{8906029, 7883913, fraire2021scalability} start routing data only after that data is requested, which requires extra processing time within the satellite communication system. To reduce such processing time, some works \cite{9377456, 6363993, 1264069} applied in-network caching in satellite networks. The authors in \cite{9377456} proposed a cache-enabled satellite-UAV-vehicle system for energy efficient data delivery services, and formulated the cache placement problem as an optimization problem. In \cite{6363993}, the problem of cache placement within information-centric satellite networks is investigated based on a profile of users' interests in different topics. The work in \cite{1264069} proposes a stochastic model to predict content popularity to help the satellite system feed caches in advance.  
However, these works only consider known, fixed service requests. Indeed, the optimization-based solutions in \cite{8906029, 7883913, fraire2021scalability, 9377456, 6363993, 1264069} may not be suitable for the design of contact plan or cache placement in real-world, highly dynamic satellite networks with dynamic, unpredictable user requests.   

 {More recently, there has been significant interest in realizing dynamic resilient satellite networking by employing machine learning tools \cite{liu2020analysis, 8353863, 8713802, pacheco2020framework, na2018distributed, 8910638, 9457160}. In particular, the work in \cite{liu2020analysis} employs a machine learning method to predict the future service needs in satellite communication networks. The authors in \cite{8353863} develop a centralized machine learning algorithm that enables real-time estimation of the environment, specifically, the ever-changing rain intensity on broadband satellite communication links. In \cite{8713802}, a deep reinforcement learning (DRL) solution is developed for intelligent satellite communications within the national aeronautics and space administration (NASA)'s space communication and navigation testbed. Yet when managing operations on different satellites, a centralized solution such as the one in \cite{8713802} can cause significant communication overhead, especially for high latency satellite systems.  Thus, distributed solutions are more desirable \cite{9562559}. In this regard, the authors in \cite{pacheco2020framework} develop a distributed  hierarchical classification solution using deep learning to intelligently classify the dynamic Internet traffic flows on satellites with low overhead.  The work in \cite{na2018distributed} employs a distributed extreme learning machine solution to route data among LEO satellites based on the forecasted dynamic traffic density. In \cite{8910638}, the authors develop a multi agent reinforcement learning (MARL) solution that enables multiple satellites to cooperatively manage their spectrum use. 
 
 Even though the machine learning based solutions developed in \cite{8353863, liu2020analysis, 8713802, pacheco2020framework, na2018distributed, 8910638} are capable of accomplishing certain networking tasks with unknown service needs or unknown communication environments, the developed solutions are mostly designed in a way to overfit to the target tasks. In particular, in the learning solutions of \cite{8353863, liu2020analysis, 8713802, pacheco2020framework, na2018distributed, 8910638}, the models are trained to serve specific data needs within specific communication environment. When serving new and unseen data needs, the models must be retrained, which incurs excessive computational costs.  As the data needs change constantly in practical applications, the satellite networks must spent a great amount of time and energy on training the machine learning based networking solutions. 
 
To reduce such cost for dynamic resilient operations, the notion of \emph{meta learning} is introduced to generalize the machine learning solutions for a family of tasks that specified by a certain distribution \cite{vanschoren2018meta}. In particular, a meta-learning model is trained over sample tasks from this distribution that serves as the initial during the regular model training for a specific unseen task, such that the regular training can be accomplished within a small number of epochs\cite{9457160, nichol2018first, vanschoren2018meta}.  In this paper, we will employ such meta learning technique to reduce the training cost. However, the satellite communication system is deployed to serve users around the globe, whose needs are different, dynamic and may follow diverse distributions, which motivates generalized meta-learning solutions that can efficiently obtain meta initials for a large number of tasks distributions. We will develop such a cost effective solution for the satellite networks.

\subsection{Contributions}
The main contribution of this paper is a novel meta reinforcement learning framework for dynamic resilient pre-storing and routing design in cube satellite networks. In particular, we consider a cube satellite communication system in which the satellites must deliver data service to ground users with dynamic and unpredictable needs. Within this system, the ground gateways will pre-store some selected data to the satellites, either directly, or by routing through the neighboring satellites. The satellites can, then, deliver data service to the target users once the data is requested.  
To achieve high pre-store hit rate, a policy needs to be designed that determines what data should be stored on each satellite and how the data is routed from the gateway to the destination satellite using the limited contact chances in the system. 
We formulate this pre-storing and routing design problem as a decentralized Markov decision process (Dec-MDP), and seek to find the optimal pre-storing and routing strategy which maximizes the fraction of user requests being directly served with the pre-stored data.

We then propose a distributed distribution-robust meta reinforcement learning (D$^2$-RMRL)  algorithm that is shown to reach a high pre-store hit rate of dynamic service needs, with low communication overhead and computation cost. In particular, to reduces the communication overhead in distributed learning, we use the value decomposition technique to reinforce the team benefit on each data flow without exchanging their action choices and environmental observation. To reduce the learning cost in the system, we use the meta training mechanism to initialize the learning procedure based on the prior information on possible data needs at different service occasions. Moreover, we use the pre-training technique to implement a shortened meta training procedure that obtains the meta initial models for a large number of service distributions. 

Simulation results show that the proposed value decomposition technique can lead to a $31.8\%$ improvement of the pre-store rate achieved by the distributed reinforcement learning algorithm. The meta learning technique can find learning initials that results in a convergence speedup by $40.7\%$. Furthermore, the meta learning procedure is shortened by up to $43.7\%$ with the proposed meta pre-training scheme under multiple service distributions. 

The remainder of this paper is organized as follows. The system model and problem formulation are described in Section \uppercase\expandafter{\romannumeral2}. In Section \uppercase\expandafter{\romannumeral3}, the proposed algorithm, including the value-decomposition-based actor-critic reinforce learning algorithm, the meta training algorithm, and the pre-training algorithm are presented. In Section \uppercase\expandafter{\romannumeral4}, simulation results are analyzed. Finally, conclusions are drawn in Section \uppercase\expandafter{\romannumeral5}.}

\section{System Model and Problem Formulation}

\begin{figure}
\setlength{\belowcaptionskip}{-16pt}
  \centering
  \includegraphics[width=9 cm]{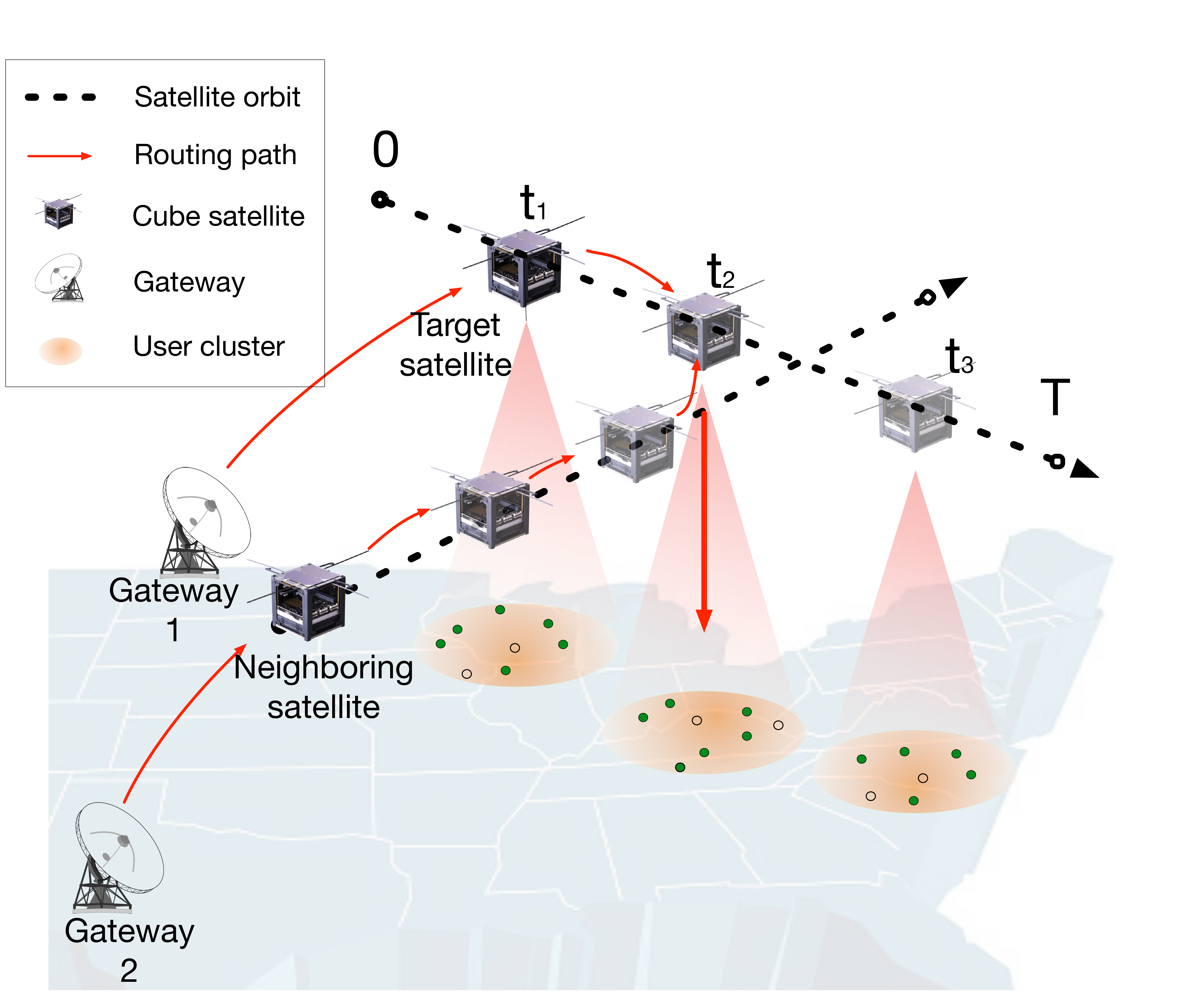}
  \caption{\footnotesize{Topology of a cube satellite network (CSN).}}
  \label{Fig. 02}
  \centering
\end{figure}


Consider a cube satellite network (CSN) that consists of $N_{\textrm{S}}$ LEO cube satellites, and $N_{\textrm{G}}$ distributed ground gateways.  At each time slot, the satellites serve $N_{\textrm{U}}$ user clusters, each of which represents a group of users that falls within the pre-store hits of a satellite, as shown in Fig. \ref{Fig. 02}. 
There are totally $N_{\textrm{F}}$ content files available in the system. At each time slot, each user cluster requests its associated satellite to deliver some content files. At this point, if the requested files are stored on the associated satellite already, users in the cluster can directly download them. Otherwise, the satellite must seek the requested files from neighboring satellites or ground gateways, such that the content requests can be served, although with higher latency.  Thus, the CSN system will pre-store the content files of interest on the satellites to serve user requests with minimum latency.

In the considered system, the gateways determine the content files that should be pre-stored on the satellites, and optimize routing path of these content files. Note that, when pre-storing content files to the satellites, the gateways do not know which files will be requested by the user clusters.  This is not only because that the user requests happen in the future, but also the users' interests on content files are highly dynamic (i.e., the interests follow unknown distributions, which also change over time and user locations\cite{chen2016caching}). On the other hand, due to the limited storage capacity, the satellites cannot store all content files that are of probable interest to the users. Thus, the gateways must selectively pre-store content files on satellites. Yet, the orbiting satellites may not be able to receive all content files of interests either, as they only have limited chances to communicate with the gateways. Then, the gateways should store some content files on the neighboring satellites that can be offloaded to the target satellites as in Fig. \ref{Fig. 02}. That is, when there are enough chances for the gateways and the target satellite to communicate, the gateways can directly store all files of interest on the target satellites. Otherwise, the gateways store some of the content files of interest on the target satellite, and the rest on the neighboring satellites while specifying how they are routed to the target satellite. In summary, the gateways in the CSN system determine how the content files are pre-stored on and routed to the satellites, based on the time-evolving CSN topology, network resource limitations, and user needs. Next, the transmission opportunities with storage limitations in the CSN system are modeled as an time-unrolled directed graph. Then, the problem formulation is given.

\subsection{Data Transmission Graph}
\begin{figure}
\setlength{\belowcaptionskip}{-16pt}
  \centering
  \includegraphics[width=9 cm]{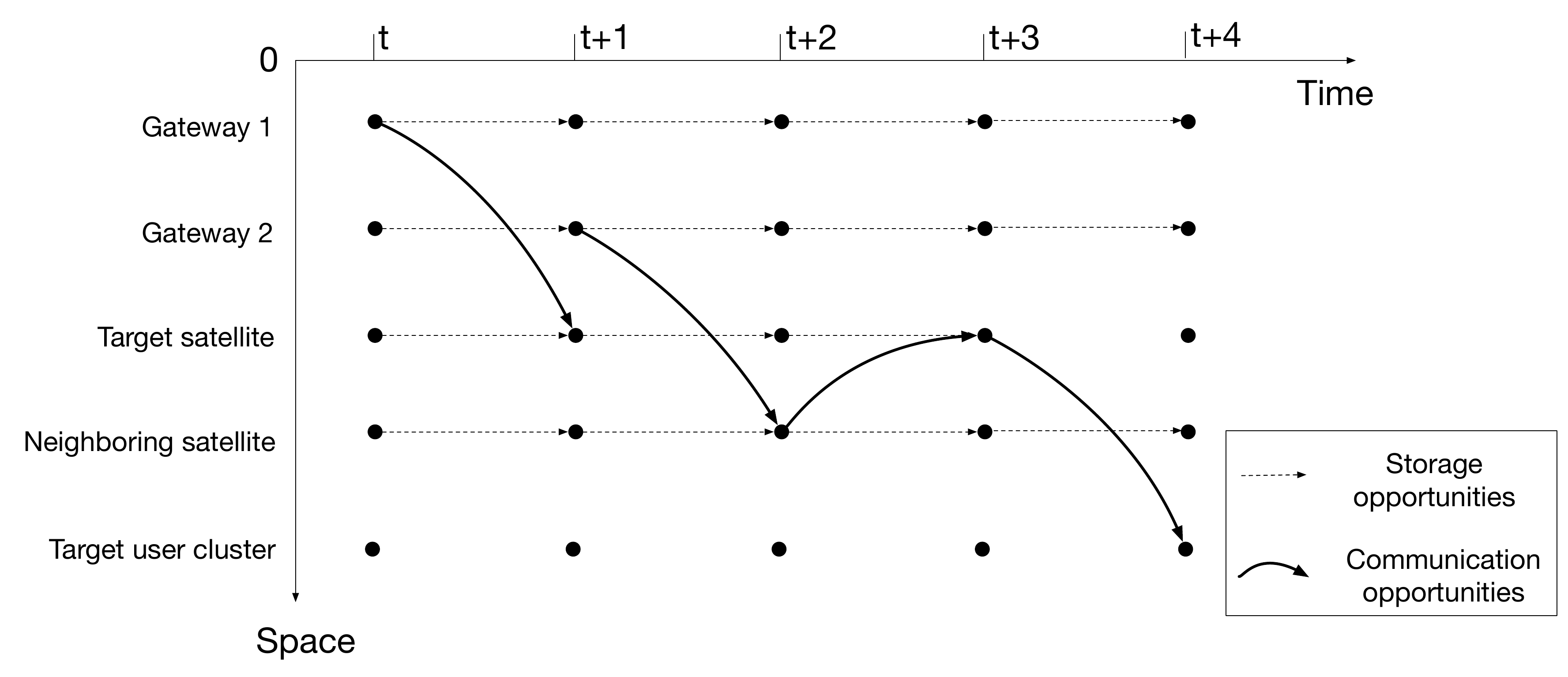}
   \vspace{-0.3cm}
  \caption{\footnotesize{Time-unrolled graph for modeling transmissions and storage opportunities in the CSN.}}
  \label{Fig. graph}
  \centering
  \vspace{-0.2cm}
\end{figure}
We use a time-unrolled graph to characterize the transmission graph evolution and storage limitations within the CSN system. The $t$-th layer of this graph represents the transmission opportunities at time slot $t$. 
The vertices at the $t$-th layer, i.e., $\mathcal{U}\left(t\right)\cup\mathcal{S}\left(t\right)\cup \mathcal{G}\left(t\right)$, correspond to the replicas of user clusters, satellites and gateways at time slot $t$. In particular, $u\in\mathcal{U}\left(t\right)$, $s\in\mathcal{S}\left(t\right)$ and $g\in\mathcal{G}\left(t\right)$ denote, respectively, a user cluster, satellite and gateway in the CSN system at time slot $t$.  

An edge in the graph connects a node at layer $t$ to another node at layer $t+1$, representing either transmitting or storing a file during time slot $t$. In particular, the set of edges $\mathcal{E}^{\textrm{G}}\!\left(t\right)=\left\{g \in \mathcal{G}\left(t\right) \to g \in \mathcal{G}\left(t+1\right)\right\}$ implies that a file that resides in a gateway $g$ at the beginning of slot $t$, will remain in this gateway during slot $t$, and therefore, it remains in the same gateway $g$, at the beginning of slot $t+1$. Similarly, the set of edges $\mathcal{E}^{\textrm{S}}\!\left(t\right)=\left\{s \in \mathcal{S}\left(t\right) \to s \in \mathcal{S}\left(t+1\right)\right\}$ denotes that a file stays in a satellite $s$ during time slot $t$. 
There are three types of file transmissions during time slot $t$:  the set of edges $\mathcal{E}^{\textrm{GS}}\!\left(t\right)=\left\{g \in \mathcal{G}\left(t\right) \to s \in \mathcal{S}\left(t+1\right)\right\}$ denotes that a file is transmitted from gateway $g$ to satellite $s$; the set of edges $\mathcal{E}^{\textrm{SS}}\!\left(t\right)\!=\!\left\{s \in \mathcal{S}\left(t\right) \to s' \in \mathcal{S}\left(t+1\right)\right\}$ models that a file is transmitted from satellite $s$ to a neighboring satellite $s'$; while the set $\mathcal{E}^{\textrm{SU}}\!\left(t\right)\!=\!\left\{ s\in\mathcal{S}\left(t\right)\! \to u\in\!\mathcal{U}\left(t+1\right)\right\}$ models that a file is transmitted from satellite $s$ to user cluster $u$. Note that, the communication opportunities in the CSN system are constrained by the communication range, and transponders on the satellites. Thus, a device can only communicate with a target satellite when it falls within this satellite's communication range, and the target satellite is not transmitting or receiving content files at current time slot $t$. In the considered CSN system, the cube satellite constellation can provide globally seamless coverage, which means that, at any time slot $t$, for any satellite $s \in {\mathcal{S}}\left(t\right)$, there always exists one and only one incoming edge $g \to s$ from a gateway $g\in\mathcal{G}\left(t-1\right)$, as well as one and only one outgoing edge $s \to u$ to an user cluster $u\in\mathcal{U}\left(t+1\right)$.  An inter-satellite communication edge $s \to s'$ exists only when the two satellites, $s\in\mathcal{S}\left(t\right)$ and $s'\in\mathcal{S}\left(t+1\right)$, orbit within each other's communication range. There is no edge linking two different gateways, two different user clusters, or a gateway and a user cluster.  Moreover, associated with each satellite $s\in\mathcal{S}\left(t\right)$ there is a variable storage capacity $\Gamma_s$, which denotes the number of files that are stored in $s$ at the current time slot. Since each satellite has limited storage, we have $0 \le \Gamma_s \le\Gamma_{\textrm{max}}$. Meanwhile, at any time slot, if a node  is the source or destination of more than one edges, only one of these edges can be active. Also, notice that the data rates of the communication edges are determined by the surrounding propagation environment\cite{tse2005fundamentals}, and we assume that a communication edge guarantees the completion of a content file transmission within one time slot.

\subsection {Pre-store Hit}
Given that the goal of the CSN system is to reduce service latency by pre-storing interesting content files on the satellites, the system performance is  evaluated by the \emph{pre-store hit}, which is defined as the number of user content requests being directly served with the files pre-stored on satellites.  In particular, a user cluster can be served with content files on satellites only if its content files of interest are already pre-stored on its associated satellite, by the time it requests these content files. For each user cluster $u$, define an $N_{\textrm{F}}\times1$ vector $\boldsymbol{x}_{u}\left(t\right)=\left[{x}^{1}_{u}\left(t\right),\ldots,{x}^{N_{\textrm{F}}}_{u}\left(t\right)\right]$, where
  \begin{equation}\label{eq:requests} 
\setlength{\belowdisplayskip}{1 pt}
\begin{split}
&{x}^{f}_{u}\left(t\right)=\\
&\;\;\;\left\{ {\begin{array}{*{20}{c}}
{1}, \;\;\textrm{if}\; u\; \textrm{requests file}\;  f\; \textrm{at the beginning of time slot}\; t\\
\!\!\!\!\!\!\!\!\!\!\!\!\!\!\!\!\!\!\!\!\!\!\!\!\!\!\!\!\!\!\!\!\!\!\!\!\!\!\!\!\!\!\!\!\!\!\!\!\!\!\!\!\!\!\!\!\!\!\!\!\!\!\!\!\!\!\!\!\!\!\!\!\!\!\!\!\!\!\!\!\!\!\!\!\!\!\!\!\!\!\!\!\!\!\!\!\!{0},\;\;\textrm{otherwise}
\end{array}} \right..
\end{split}
\end{equation}  
We assume that if a file $f$ is transmitted to a satellite $s$ in time slot $t$, then it is stored in $s$, at the beginning of time slot $t+1$. For each satellite $s$, define an $N_{\textrm{F}}\times1$ vector $\boldsymbol{y}_{s}=\left[{y}^{1}_{s},\ldots,{y}^{N_{\textrm{F}}}_{s}\right]$, where $y^f_s$ denotes the ``age'' of file $f$ at $s$, i.e., 
  \begin{equation}\label{eq:availabilities} 
\begin{split}
&{y}^{f}_{s}= \\  
&\;\;\;\left\{ {\begin{array}{*{20}{c}}
{k}, \;\;\textrm{if}\; f \; \textrm{is stored on}\;  s\; \textrm{at the beginning of time slot}\; k\\
\!\!\!\!\!\!\!\!\!\!\!\!\!\!\!\!\!\!\!\!\!\!\!\!\!\!\!\!\!\!\!\!\!\!\!\!\!\!\!\!\!\!\!\!\!\!\!\!\!\!\!\!\!\!\!\!\!\!\!\!\!\!\!\!\!\!\!\!\!{\infty },\;\;\textrm{if}\; f \;\textrm{is not stored on}\; s 
\end{array}} \right.,
\end{split}
\end{equation} 
with $k=1, 2, ...$.
Then, the total hits at time slot $t$ is
  \begin{equation}\label{eq:utility} 
\begin{split}
&h\left(t\right)=\sum\limits^{N_{\textrm{F}}}_{f=1}\sum\limits_{u\in\mathcal{U}\left(t\right)}\sum\limits_{\substack{s\in\mathcal{S}\left(t\right), \\ s \to u\in\mathcal{E}^{\textrm{SU}}\left(t\right)}}\mathds{1}_{\left\{{x}^{f}_{u}\left(t\right) = 1,  {y}^{f}_{s}\neq \infty \right\}},
\end{split}
\end{equation}  
with 
$\mathds{1}_{\chi} = 1$ when $\chi$ is true, otherwise, $\mathds{1}_{\chi} = 0$. Thus, $\mathds{1}_{\left\{{x}^{f}_{u}\left(t\right) =1, {y}^{f}_{s} \neq \infty \right\}}=1$ represents that user cluster $u$ can directly download content file $f$ from satellite $s$, at slot $t$. 
The pre-storing and routing process in the considered CSN system aims at the highest number of pre-store hits, under the network topology evolution and resource limitations.  This pre-storing and routing process is modeled in the next subsection. 

 \subsection {MDP Modeling for a Single Content File}
In our system model, the gateways make plans on how each content file is pre-stored and routed within a CSN system. In particular, only the content files that are deemed of interest to the ground users should be pre-stored on the target satellites, or neighboring satellites, using the limited communication and storage opportunities in the system. However, the resulting pre-store hits is partially determined by the gateways' decisions, and partially determined by the dynamic, and unforeseeable user requests in the system. The routing process of a given content file $f$ can be formulated as a Markov decision process (MDP) $\left\langle {{\Omega}^f, \mathcal{A}^f, R^f} \right\rangle$\cite{puterman2014markov}. We next describe the state space $\Omega^f$, the action space $\mathcal{A}^f$ and the reward function $R^f$. 


\subsubsection{State}  $\Omega^f$ is the state space of content file $f$. The state is a pair ${\sigma}^f\left(t\right)=\left[v, x\right]$, where for a regular state  $v \in \mathcal{G}\left(t\right)\bigcup\mathcal{S}\left(t\right)$, and ${x}\in\left\{0,1,\emptyset\right\}$.  Specifically, $v$ denotes the location of file $f$ at time $t$. If $v=g \in {\mathcal{G}}\left(t\right)$, i.e., file $f$ is in gateway $g$, then $x = \emptyset$. 
On the other hand, if $v=s \in {\mathcal{S}}\left(t\right)$, i.e., file $f$ is in satellite $s$ at time $t$, then $x = {x}^f_u\left(t\right)$ indicates whether or not file $f$ is requested by the user cluster $u$ currently connected to $s$, i.e., $s\to u\in\mathcal{E}^{\textrm{SU}}\left(t\right)$.  Moreover, there is an initial state denoted as $\left[ {\sf I}, \emptyset\right]$ and a terminal state denoted as $\left[{\sf T}, \emptyset\right]$. A file always starts from the initial state and stays there until it is picked by a gateway, and it terminates when the satellite that stores it decides to discard it. 

\subsubsection{Action}  $\mathcal{A}^f$ is the action space of content file $f$. The possible actions ${a}^f\left(t\right)$ corresponding to the state ${\sigma}^f\left(t\right)$, and the resulting state ${\sigma}^f\left(t+1\right)$ are as follows; and we denote the corresponding probabilistic policy as $\pi^f \left( a^f\left(t\right) \left. | \sigma^f\left(t\right) \right. \right)$. 
\begin{itemize}
\item For the initial state ${\sigma}^f\left(t\right) = \left[ {\sf I}, \emptyset\right]$, there are two possible actions:
\begin{enumerate}[label=(\alph*)]
\item  ${a}^f\left(t\right)= {\sf I} \to {\sf I}$, i.e., the file stays at the initial state, and hence ${\sigma}^f\left(t+1\right) = \left[ {\sf I}, \emptyset\right]$. 
\item ${a}^f\left(t\right) = {\sf I} \to g \in {\mathcal{G}}\left(t\right)$, i.e., the file is picked by gateway $g$, and hence ${\sigma}^f\left(t+1\right) = \left[ g, \emptyset\right]$.
\end{enumerate}

\item For a regular state ${\sigma}^f\left(t\right)=\left[g, \emptyset \right]$, there are two possible actions:
\begin{enumerate}[label=(\alph*)]
\item ${a}^f\left(t\right) = g\to g\in\mathcal{E}^{\textrm{G}}\left(t\right)$, i.e., file $f$ stays in gateway $g$, leading to the next state ${\sigma}^f\left(t+1\right)=\left[g, \emptyset \right]$. 

\item ${a}^f\left(t\right) = g\to s\in\mathcal{E}^{\textrm{GS}}\left(t\right)$, i.e., file $f$ is transmitted from gateway $g$ to satellite $s$, leading to the next state ${\sigma}^f\left(t+1\right)=\left[s, x^f_{u}\left(t+1\right) \right]$, with $s\to u \in \mathcal{G}^{\textrm{SU}}\left(t+1\right)$. As part of this action, we record the time file $f$ is stored in satellite $s$ with ${y}^{f}_{s}\leftarrow t+1$, and update the occupancy of $s$ as $\Gamma_s \leftarrow  \Gamma_s+1$. Now, if $\Gamma_s = \Gamma_{\textrm{max}}$, i.e., the storage capacity is reached, then the oldest file on $s$ is discarded, i.e., $y_s^{\hat f}  \leftarrow \infty$, where $\hat f = \arg\min_{f\in\mathcal{F}} y_s^f$, and $\Gamma_s \leftarrow \Gamma_s - 1$. Moreover, the file $\hat f$ reaches the terminal state, i.e., $\sigma^{\hat f} \left(t+1\right) = \left[{\sf T}, \emptyset\right]$. Hence, the action ${a}^f\left(t\right)$ of file $f$ can affect the state $\sigma^{\hat f} \left(t+1\right)$ of another file $\hat f$. 

\end{enumerate}
\item For a regular state ${\sigma}^f\left(t\right)=\left[s, x\right]$, there are three possible actions:
 \begin{enumerate}[label=(\alph*)]
\item  If $x= 1$, i.e., file $f$ is requested by user cluster $u$ that is connected to $s$, then ${a}^f\left(t\right) = s\to u\in\mathcal{E}^{\textrm{SU}}\left(t\right)$, i.e., file $f$ is transmitted from satellite $s$ to user cluster $u$, leading to the next state ${\sigma}^f\left(t+1\right)=\left[s, {x}^f_u\left(t+1\right) \right]$, with $s\to u\in\mathcal{E}^{\textrm{SU}}\left(t+1\right)$. 

\item If $x= 0$, then ${a}^f\left(t\right)$ can be $s\to s\in\mathcal{E}^{\textrm{S}}\left(t\right)$, i.e., file $f$ stays in satellite $s$, leading to the next state ${\sigma}^f\left(t+1\right)=\left[s, {x}^f_u\left(t+1\right) \right]$, with $s\to u\in\mathcal{E}^{\textrm{SU}}\left(t+1\right)$. 

\item Or if $x= 0$, ${a}^f\left(t\right)$ can also be $s\to s'\in\mathcal{E}^{\textrm{SS}}\left(t\right)$, i.e., file $f$  is transmitted from satellite $s$ to its neighboring satellite $s'$, leading to the next state ${\sigma}^f\left(t+1\right)=\left[s', {x}^f_u\left(t+1\right) \right]$, with $s'\to u\in\mathcal{E}^{\textrm{SU}}\left(t+1\right)$.  As part of this action, we record the time file $f$ is stored in satellite $s'$ with ${y}^{f}_{s'}\leftarrow t+1$, and set ${y}^{f}_{s}\leftarrow \infty$, since file $f$ is moved out of satellite $s$. 
We update the occupancy of $s$ as $\Gamma_s \leftarrow  \Gamma_s-1$, and the occupancy of $s'$ as $\Gamma_{s'} \leftarrow  \Gamma_{s'}+1$. Now, if $\Gamma_{s'} = \Gamma_{\textrm{max}}$, i.e., the storage capacity is reached, then the oldest file on $s'$ is discarded, i.e., $y_{s'}^{\hat f}  \leftarrow \infty$, where $\hat f = \arg\min_{f\in\mathcal{F}} y_{s'}^f$, and $\Gamma_{s'} \leftarrow \Gamma_{s'} - 1$. Moreover, the file $\hat f$ reaches the terminal state, i.e., $\sigma^{\hat f} \left(t+1\right) = \left[{\sf T}, \emptyset\right]$. 
\end{enumerate}

\item For the terminal state ${\sigma}^f\left(t\right) = \left[{\sf T}, \emptyset\right]$, no action is needed and the state remains terminal. 

\end{itemize}

\subsubsection{Reward}  The reward function $R^f\left(t\right)=R^f\left({a}^f\left(t\right), {\sigma}^f\left(t\right)\right)$ evaluates action choices ${a}^f\left(t\right)\in\mathcal{A}^f$ at different states ${\sigma}^f\left(t\right)\in\Omega^f$. Since the system performance metric is the total number of hits in (\ref{eq:utility}), we get a unit reward at time $t$ corresponding to a hit if and only if the action is a file transmission from a satellite to its user cluster, i.e.,  $R^f\left(t\right)=\mathds{1}_{\left\{{\sigma}^f\left(t\right) = \left[s, 1\right] \right\}}$. 

 \subsection {Problem Formulation}
The routing process in the CSN system considers multiple content files. The routing decisions on one file $f$ can directly affect those on some other content files, since the files compete for the communication or storage resources of the satellites. Thus, the routing of multiple content files should be cooperatively arranged by the gateways.  
To model these interdependent decision making processes, we formulate a Dec-MDP defined by $\left\langle {\mathcal{F}, \Omega, \mathcal{A}, R} \right\rangle$\cite{puterman2014markov}, where 
\begin{itemize}
\item $\mathcal{F}=\left\{1, ..., N_{\textrm{F}}\right\}$ is the set of content files to be routed in the CSN system. 

\item $\Omega=\Omega^1\times \ldots \times\Omega^{N_{\textrm{F}}}$ is the state space. The state of the multiple content routing process at time slot $t$ is captured as $\boldsymbol{\sigma}\left(t\right)=\left({\sigma}^1\left(t\right),\ldots, {\sigma}^{N_{\textrm{F}}}\left(t\right)\right)$. 

\item $\mathcal{A}\subset\mathcal{A}^1\times \ldots \times\mathcal{A}^{N_{\textrm{F}}}$ is the action space. The action of the multiple content routing process at time slot $t$ is captured as $\boldsymbol{a}\left(t\right)=\left({a}^1\left(t\right),\ldots, {a}^{N_{\textrm{F}}}\left(t\right)\right)$. Note that, due to the possible conflicts in the system, e.g., multiple routing processes compete for one communication or storage chance, the action space $\mathcal{A}$ is a strict subset of $\mathcal{A}^1\times \ldots \times\mathcal{A}^{N_{\textrm{F}}}$, and captures only the viable action choices under resource limitations. In particular, at time slot $t$, if every file $f$ follows its own local policy $\pi^f \left( a^f\left(t\right) \left. | \sigma^f\left(t\right) \right. \right)$, then for any give file $f$, its action $a^f\left(t\right) \in {\mathcal{A}}^f$ might be in conflict with the action $a^{f'}\left(t\right) \in {{\mathcal{A}}}^{f'}$ of another file $f'$ based on its local policy $\pi^{f'} \left( a^f\left(t\right) \left. | \sigma^f\left(t\right) \right. \right)$, as follows


\begin{itemize}
\item For the initial state ${\sigma}^f\left(t\right) = \left[ {\sf I}, \emptyset\right]$, the two possible actions ${\sf I} \to {\sf I}$ and ${\sf I}\to g$ are not in conflict with the action of any other file.

\item For a regular state ${\sigma}^f\left(t\right)=\left[g, \emptyset \right]$, 
\begin{enumerate}[label=(\alph*)]
\item  Action ${a}^f\left(t\right) = g\to g\in\mathcal{E}^{\textrm{G}}\left(t\right)$ is not in conflict with others. 

\item Action ${a}^f\left(t\right) = g\to s\in\mathcal{E}^{\textrm{GS}}\left(t\right)$ is in conflict with actions ${a}^{f'}\left(t\right)$ of another file $f'$ that take gateway $g$ or satellite $s$ as the transmitter or receiver, i.e.,
  \begin{equation}\label{eq:conflict1} 
\begin{split}
&{a}^{f'}\!\!\!\left(t\right)\!=\!
\left\{ {\begin{array}{*{20}{c}}
{g\to s'\in\mathcal{E}^{\textrm{GS}}\left(t\right)}, \;\;\textrm{if}\; {\sigma}^{f'}\left(t\right)=\left[g, \emptyset\right]\\
{g'\to s\in\mathcal{E}^{\textrm{GS}}\left(t\right)},\;\;\textrm{if}\; {\sigma}^{f'}\left(t\right)=\left[g', \emptyset\right]\\
{s'\to s\in\mathcal{E}^{\textrm{SS}}\left(t\right)}, \;\;\textrm{if} \; {\sigma}^{f'}\left(t\right)=\left[s', 0\right]\\
{s\to s'\in\mathcal{E}^{\textrm{SS}}\left(t\right)}, \;\;\textrm{if} \; {\sigma}^{f'}\left(t\right)=\left[s, 0\right]\\
{s\to u\in\mathcal{E}^{\textrm{SU}}\left(t\right)}, \;\;\textrm{if}\; {\sigma}^{f'}\left(t\right)=\left[s, 1\right]
\end{array}} \right.,
\end{split}
\end{equation} 
for all $f'\in\mathcal{F}\backslash \left\{f\right\}$. 

\end{enumerate}
\item For a regular state ${\sigma}^f\left(t\right)=\left[s, 1\right]$, the only possible ${a}^f\left(t\right) = s\to u\in\mathcal{E}^{\textrm{SU}}\left(t\right)$ is in conflict with the following actions:
  \begin{equation}\label{eq:conflict2} 
\begin{split}
&{a}^{f'}\!\left(t\right)=
\left\{ {\begin{array}{*{20}{c}}
\!{g\to s\in\mathcal{E}^{\textrm{GS}}\left(t\right)}, \;\;\textrm{if}\; {\sigma}^{f'}\left(t\right)=\left[g, \emptyset\right]\\
{s'\to s\in\mathcal{E}^{\textrm{SS}}\left(t\right)},\;\;\textrm{if}\; {\sigma}^{f'}\left(t\right)=\left[s', \emptyset\right]\\
{s\to s'\in\mathcal{E}^{\textrm{SS}}\left(t\right)}, \;\;\textrm{if} \; {\sigma}^{f'}\left(t\right)=\left[s, 0\right]\\
{s\to u\in\mathcal{E}^{\textrm{SU}}\left(t\right)}, \;\;\textrm{if} \; {\sigma}^{f'}\left(t\right)=\left[s, 1\right]
\end{array}} \right.,
\end{split}
\end{equation} 
for all $f'\in\mathcal{F}\backslash\left\{f\right\}$.  

\item For a regular state ${\sigma}^f\left(t\right)=\left[s, 0\right]$, 

\begin{enumerate}[label=(\alph*)]
\item  Action ${a}^f\left(t\right)=s\to s\in\mathcal{E}^{\textrm{S}}\left(t\right)$ is not in conflict with others. 

\item Action ${a}^f\left(t\right) = s\to s'\in\mathcal{E}^{\textrm{SS}}\left(t\right)$, is in conflict with the following actions:
  \begin{equation}\label{eq:conflict1} 
\begin{split}
&{a}^{f'}\!\!\!\!\left(t\right)\!=\!
\left\{ {\begin{array}{*{20}{c}}
{g\to s\in\mathcal{E}^{\textrm{GS}}\left(t\right)}, \;\;\textrm{if}\; {\sigma}^{f'}\left(t\right)=\left[g, \emptyset\right]\\
{g\to s'\in\mathcal{E}^{\textrm{GS}}\left(t\right)}, \;\;\textrm{if}\; {\sigma}^{f'}\left(t\right)=\left[g, \emptyset\right]\\
{s''\to s\in\mathcal{E}^{\textrm{SS}}\left(t\right)},\;\;\textrm{if}\; {\sigma}^{f'}\left(t\right)=\left[s'', 0\right]\\
{s''\to s'\in\mathcal{E}^{\textrm{SS}}\left(t\right)}, \;\;\textrm{if} \; {\sigma}^{f'}\left(t\right)=\left[s'', 0\right]\\
{s\to s''\in\mathcal{E}^{\textrm{SS}}\left(t\right)}, \;\;\textrm{if} \; {\sigma}^{f'}\left(t\right)=\left[s, 0\right]\\
{s'\to s''\in\mathcal{E}^{\textrm{SS}}\left(t\right)}, \;\;\textrm{if}\; {\sigma}^{f'}\left(t\right)=\left[s', 0\right]\\
{s\to u\in\mathcal{E}^{\textrm{SU}}\left(t\right)}, \;\;\textrm{if} \; {\sigma}^{f'}\left(t\right)=\left[s, 1\right]\\
{s'\to u\in\mathcal{E}^{\textrm{SU}}\left(t\right)}, \;\;\textrm{if} \; {\sigma}^{f'}\left(t\right)=\left[s', 1\right]
\end{array}} \right.,
\end{split}
\end{equation} 
for all $f'\in\mathcal{F}/\left\{f\right\}$.  


\end{enumerate}


\end{itemize}


\item ${\small {R\left(\boldsymbol{a}\left(t\right), \boldsymbol{\sigma}\left(t\right)\right)\!\!=\!\! \sum_{f\in\mathcal{F}}R^f\left(t\right)= \sum_{f\in\mathcal{F}} \mathds{1}_ {\left\{\sigma^f\left(t\right) = \left[s \in \mathcal{S}\left(t\right), 1\right]\right\}}}}$ is the reward function that evaluates the action choice $\boldsymbol{a}\left(t\right)$ at state $\boldsymbol{\sigma}\left(t\right)$, which simply counts the number of hits at time slot $t$.  
\end{itemize}

Here, our goal is to find a global policy $\pi\left(\boldsymbol{a}\left(t\right)\left|\boldsymbol{\sigma}\left(t\right)\right.\right)$ that provides a probabilistic mapping from each joint state $\boldsymbol{\sigma}\left(t\right)$ to the corresponding joint action $\boldsymbol{a}\left(t\right)$ that is conflict-free. However, since the number of content files $N_{\textrm{F}}$ is typically large, the underlying MDP is very high-dimensional and hard to solve or implement in practice. Therefore, in this work, we target at distributed probabilistic policies together with a conflict resolution scheme. In particular, associated with each file $f$ there is a local policy $\pi^f\left(a^f\left(t\right)\left.|\sigma^f\left(t\right)\right.\right)$ that is a probabilistic mapping from a local state $\sigma^f\left(t\right)$ to the corresponding action $a^f\left(t\right)\in\mathcal{A}^f$. Note that, such distributed policies may result in actions $\boldsymbol{a}\left(t\right)=\left[a^1\left(t\right),\ldots, a^{N_{\textrm{F}}}\left(t\right)\right]$ that are in conflict, which must be resolved when being implemented in the considered CSN system. Since our  goal is to maximize the pre-store hits, we should retain as many ``$s\to u$'' actions, and disable the routing actions that are in conflict with ``$s\to u$''. 
Hence, given the probabilistic local policies $\pi^f\left(a^f\left(t\right)\left.|\sigma^f\left(t\right)\right.\right)$, $f=1,\ldots, N_{\textrm{F}}$ and the states $\boldsymbol{\sigma}\left(t\right)=\left[\sigma^1\left(t\right),\ldots, \sigma^{N_{\textrm{F}}}\left(t\right)\right]$, the procedure for resolving the conflicts is summarized in Algorithm \ref{alg:conflicts}. 
The procedure incrementally form the conflict-free action set $\mathcal{T}$. First, in Lines 1-9, $\mathcal{T}$ consists of all $s\to u$ actions and here the only possible conflict is that one satellite may be scheduled to transmit more than one file to the user cluster it connects to. In that case, we allow only one file to be transmitted and let the other files remain on the satellite. Then in Lines 10-15, we examine each action not in $\mathcal{T}$. If it is in conflict with any action in $\mathcal{T}$, it must be either $g\to s$  corresponding to Eq. (4), or $s\to s'$ corresponding to Eq. (6), and we simply keep the in-conflict files on their current location, i.e., gateway $g$, or satellite $s$. 
\begin{remark}
{Note that from the view point of MDP, in our approach, the global  policy is decomposed into independent local policies, i.e., $\pi\left(\boldsymbol{a}\left(t\right)\left.|\boldsymbol{\sigma}\left(t\right)\right.\right)=\prod\limits_{f=1}^{N_{\textrm{F}}} \pi^f\left(a^f\left(t\right)\left.|\sigma^f\left(t\right)\right.\right)$. Then for each file $f$, the conflict-resolving Algorithm 1 plays the role of the environment that determines the next state given the current state and action, i.e., $p\left(\sigma^f\left(t+1\right)\left.|\sigma^f\left(t\right), a^f\left(t\right)\right.\right)$. }
\end{remark}

Our problem is then to design the local probabilistic policies such that when employed by the above distributed routing algorithm, the maximum pre-store hit rate can be achieved. 

{\color{black}  \begin{algorithm}[t]\small

\caption{Procedure for resolving conflicts among actions produced by local policies.}   

\label{alg:conflicts}   
\setlength{\abovecaptionskip}{-40pt} 
\setlength{\belowcaptionskip}{-40pt}
\begin{algorithmic} [1] 
\REQUIRE Joint state $\boldsymbol{\sigma}\left(t\right)=\left[\boldsymbol{\sigma}_1\left(t\right),\ldots, \boldsymbol{\sigma}_{N_{\textrm{U}}}\left(t\right)\right]$, local policies $\pi^f\left(a^f\left(t\right)\left.|\sigma^f\left(t\right)\right.\right)$, $f=1, \ldots, N_{\textrm{F}}.$ \\ 
\vspace{2pt}  
\ENSURE $\mathcal{T} =\emptyset $.\\
\vspace{2pt}  
\FOR {$f=1,\ldots, N_{\textrm{F}}$} 
\vspace{2pt}  
\STATE Generate local actions $a^f\left(t\right)$ according to $\pi^f\left(a^f\left(t\right)\left.|\sigma^f\left(t\right)\right.\right)$. 
\IF {$a^f\left(t\right)=s\to u$} 
\vspace{2pt}
\STATE $\mathcal{T}=\mathcal{T}\bigcup \left\{f\right\}$. 
\vspace{2pt}
\ENDIF
\vspace{2pt} 
\STATE Check all files in $\mathcal{T}$:
\vspace{2pt} 
\IF {more than one file $f_1, \ldots, f_k$ are located on the same satellite $s$} 
\vspace{2pt}
\STATE Randomly select one file $f_j$ to be moved to $u$, and the rest stay on $s$, , i.e., $\sigma^{f_j} \left(t+1\right) = u, 
\sigma^{f_{j'}}\left(t+1\right)=s$, $j'\in \left\{1, ..., k\right\} \backslash \left\{j\right\}$. 
\ENDIF
\vspace{2pt}  
\FOR {each file $f \notin \mathcal{T}$} 
\vspace{2pt} 
\IF {$a^f\left(t\right)$ is in conflict with any action in $\mathcal{T}$} 
\vspace{2pt}
\IF {$a^f\left(t\right)=g\to s$} 
\vspace{2pt}
\STATE File $f$ stays on gateway $g$, i.e. ${\sigma}^f\left(t+1\right)=g$. 
\vspace{2pt}
\ELSIF {$a^f\left(t\right)=s\to s'$} 
\vspace{2pt}
\STATE File $f$ stays on satellite $s$, i.e. ${\sigma}^f\left(t+1\right)=s$. 
\vspace{2pt}
\ENDIF
\vspace{2pt}  
\ELSE
\vspace{2pt}  
\STATE $\mathcal{T}=\mathcal{T}\bigcup \left\{f\right\}$. 
\vspace{2pt}
\ENDIF
\vspace{2pt}  
\ENDFOR
\vspace{2pt}  
\ENDFOR
\vspace{2pt}  
\RETURN{ Next state $\boldsymbol{\sigma}\left(t+1\right)$ resulting from actions $\boldsymbol{a}\left(t\right)=\left[a^1\left(t\right),\ldots, a^{N_{\textrm{F}}}\left(t\right)\right]$,  at state $\boldsymbol{\sigma}\left(t\right)$. }
\end{algorithmic}
\end{algorithm}}

In the considered CSN system, the gateways have access to all content files in $\mathcal{F}$ as they are connected to the core network. Targeting at the highest number of pre-store hits, we solve the Dec-MDP to obtain the optimal policy for the routing of content files. Then, the gateways will transmit the files to the satellites, along with the commands on these files' routing strategies. 
However, we notice that the user requests are hard to satisfy, since 1) the user requests $\bold{X}=\left\{\left[\boldsymbol{x}_{1}\left(t\right),\ldots,\boldsymbol{x}_{N_{\textrm{U}}}\left(t\right)\right], t=1,\ldots,T\right\}$, are unknown at the time when the gateways make decisions on the routing processes; 2) the user requests $\bold{X}$ are dynamic, as they follow some unknown distributions, $p\left(\bold{X}\right)$; and 3) the user requests are unpredictable, as their distribution $p\left(\bold{X}\right)$ may vary.  Thus,  traditional MDP solutions such as decision tree search or dynamic programming can not solve this Dec-MDP. 
The reinforcement learning (RL) algorithms such as Q learning, policy gradient, and echo state networks \cite{watkins1992q, chen2019joint, sutton2000policy} can help learn routing strategies in the unknown environment, but are still not suitable for solving this high dimensional Dec-MDP. This is because for each possible realization of user requests $\bold{X}$, we need to run the RL algorithm to obtain the corresponding optimal pre-storing and routing strategy, which makes such a solution approach prohibitively complex. Thus, to solve the time sensitive pre-storing and routing problem for dynamic, unpredictable content needs, we propose a distributed distribution-robust meta reinforcement learning (D$^2$-RMRL) solution with pre-trained meta learning capability.

\section{Distribution Robust Meta Reinforcement Learning Algorithm}

{We now introduce the D$^2$-RMRL algorithm, which integrates the techniques of value decomposition \cite{sunehag2017value}, model agnostic meta-learning \cite{finn2017model}, pre-training \cite{erhan2010does}, with the actor-critic RL framework. The value decomposition effectively converts the original problem of finding the global policy $\pi\! \left(\boldsymbol a\!\left(t\right)\! \left| \boldsymbol \sigma\! \left(t\right) \right. \!\right)$ to the one of finding the set of local policies $\left\{\!\pi^f_0\!\!\left( a^f\!\left(t\right) \left|  \sigma^f \!\left(t\right) \right. \right)\!, f\!=\!1,\ldots, \! N_{\textrm{F}} \right\}$, thus significantly reduces the computational complexity. {The meta training scheme further reduces the cost of computing the optimal policy for every possible service request realization $\boldsymbol X$, by first computing a meta policy that can serve as a good initialization for service requests following distribution $p\left(\boldsymbol X\right)$, and then the optimal policy for any realization $\boldsymbol X \sim p\left(\boldsymbol X\right)$ can be obtained by a small number of gradient updates.} Finally, to address varying user request distributions, the pre-training scheme is adopted that employs the meta training procedure with a parameter transfer technique. 
In what follows, we first explain how to apply the actor-critic RL algorithm to solve the Dec-MDP using the value decomposition technique. Then we explain how this RL solution is meta-trained for faster convergence on serving different user requests. Finally, we introduce the pre-training scheme to make the performance of the proposed D$^2$-RMRL robust to different user request distributions.
\subsection{Actor Critic Method for Computing the Global Policy $\pi\left(\boldsymbol{a}\left(t\right)\left| \boldsymbol{\sigma}\left(t\right) \right.\right)$ for a Given User Request Realization $\boldsymbol X$}

{The goal of RL is to find a probabilistic policy $\pi^*$: $\Omega\!\to\!\mathcal{A}$ that provides a mapping from the CSN system's states to the actions yielding the highest cumulative discounted reward, i.e., $\pi^*=\arg\max_{\pi} \mathds{E}\left[\sum\limits_{t=1}^{T}\gamma^t  R\!\left(\boldsymbol{a}\left(t\right)\!, \boldsymbol{\sigma}\left(t\right) \right)\pi\left(\boldsymbol{a}\!\left(t\right)\left| \boldsymbol{\sigma}\!\left(t\right)\right. \right)\right] $ where $\gamma$ is the discount factor and $\left[\boldsymbol{\sigma}\left(0\right), \boldsymbol{a}\left(0\right), \boldsymbol{\sigma}\left(1\right), \boldsymbol{a}\left(1\right), \ldots \right]$ is a state-action trajectory generated by the policy $\pi$. Thus, the action choices in the CSN system must consider the instantaneous reward and the discounted future rewards. 
So, here we define the (state) value function with a given policy $\pi$ as 
  \begin{equation}\label{eq:valuedef}
\begin{split}
V^{\pi}\left( \boldsymbol{\sigma}\left(t\right) \right) &=\mathds{E}\left[\sum\limits_{\tau=t}^{T}\gamma^\tau  R\!\left(\boldsymbol{a}\left(\tau\right)\!, \boldsymbol{\sigma}\left(\tau\right) \right)\pi\left(\boldsymbol{a}\!\left(\tau\right)\left| \boldsymbol{\sigma}\!\left(\tau\right)\right. \right)\right] ,
\end{split}
\end{equation} 
which encodes the expected cumulative reward when starting in state $\boldsymbol{\sigma}\left(t\right)$ and following the policy $\pi$ thereafter. 
The optimal value over all possible policies is
 \begin{equation}\label{eq:optvaluedef}
\begin{split}
V^{*}\left( \boldsymbol{\sigma}\left(t\right) \right) &= \max\limits_{\pi:\Omega\to\mathcal{A}}V^{\pi}\left( \boldsymbol{\sigma}\left(t\right) \right). 
\end{split}
\end{equation} 
Then the optimal policy $\pi^*$ can be obtained by always picking actions $\boldsymbol{a}^*\left(t\right)$ that are greedy with respect to $V^*$, i.e., $\boldsymbol{a}^*\left(t\right)=\arg\max\limits_{\boldsymbol{a}\left(t\right) \in\mathcal{A}} R\left(\boldsymbol{a}\left(t\right) , \boldsymbol{\sigma}\left(t\right)  \right)+\gamma V^{*}\left( \boldsymbol{\sigma}\left(t+1\right)  \right)$. }
In the actor-critic RL (ACRL) method, both the policy function ${\pi}_{\theta}$ and value function $V_{{\psi}}$ are deep neural networks parameterized by ${\theta}$ and ${{\psi}}$, respectively.
Recall that under the conventional RL, for each realization of user request $\bold{X}=\left\{\left[\boldsymbol{x}_{1}\left(t\right),\ldots,\boldsymbol{x}_{N_{\textrm{U}}}\left(t\right)\right], t=1,\ldots,T\right\}$, we need to compute the corresponding policy $\pi$. 
Starting with randomly chosen parameters $\theta^{(0)}$ and $\psi^{(0)}$, the $i$-th epoch of the ACRL training algorithm consists of the following 
steps:
 
 \subsubsection{Sampling} First, using the policy ${\pi}_{{\theta}^{\left(i-1\right)}}$ from the previous epoch, the algorithm records a sampled state-action-reward trajectory and stores it as $\boldsymbol{\eta}\!=\!\left\{\boldsymbol{\sigma}\!\!\left(t\right)\!, \boldsymbol{a}\!\left(t\right)\!,  R\!\left(\boldsymbol{a}\!\left(t\right)\!, \boldsymbol{\sigma}\!\left(t\right)\right)\!, \! t\!\in\!\left[1,\ldots,T\right]\right\}$. 
 
 \subsubsection{Computing the learning error}  The temporal difference (TD) error is defined as \cite{baird1993advantage}
  \begin{equation}\label{eq:tderror}
\begin{split}
A\left(\boldsymbol{a}\left(t\right), \boldsymbol{\sigma}\left(t\right) \right) = &R\left(\boldsymbol{a}\left(t\right), \boldsymbol{\sigma}\left(t\right) \right)\\
& +\gamma V_{{\psi }^{\left(i-1\right)}}\left(\boldsymbol{\sigma}\left(t+1\right)\right)- V_{{\psi }^{\left(i-1\right)}}\left(\boldsymbol{\sigma}\left(t\right)\right),
\end{split}
\end{equation} 
which measures the difference between the achieved reward based on the samples in $\boldsymbol{\eta}$ and the estimated reward, and reveals how actions in $\boldsymbol{a}\left(t\right)$ are better than other action choices at the current state $\boldsymbol{\sigma}\left(t\right)$.  Meanwhile, the error on the routing policy is set as the expected advantage with policy ${\pi}_{{\theta}^{\left(i\right)}}$ as in
   \begin{equation}\label{eq:perror}
\begin{split}
\sum\limits_{t=1}^{T}A\left(\boldsymbol{a}\left(t\right), \boldsymbol{\sigma}\left(t\right)\right)\pi_{{\theta}^{\left(i\right)}}\left(\boldsymbol{a}\left(t\right)\left| \boldsymbol{\sigma}\left(t\right) \right.\right). 
\end{split}
\end{equation} 
 \subsubsection{Updating networks} The critic network $V_{\psi^{\left(i\right)}}$, i.e., the value function, is updated toward the opposite direction of the gradient of the squared TD error, $A^2\left(\boldsymbol{a}\left(t\right), \boldsymbol{\sigma}\left(t\right)\right)$, for accurate future reward estimation, as
 \begin{equation}\label{eq:ACvupdate}
\begin{split}
{\psi}^{\left(i\right)}&={\psi}^{\left(i-1\right)}-\alpha^{\left(i\right)}_c\nabla_{{\psi}}\sum\limits_{t=1}^{T}A^2\left(\boldsymbol{a}\left(t\right), \boldsymbol{\sigma}\left(t\right)\right) \\
&={\psi}^{\left(i-1\right)}+2\alpha^{\left(i\right)}_c   \sum\limits_{t=1}^{T} A\left(\boldsymbol{a}\left(t\right), \boldsymbol{\sigma}\left(t\right)\right)\nabla_{{\psi}}{V}_{{\psi}^{\left(i-1\right)}}\left(\boldsymbol{\sigma}\left(t\right)\right),
\end{split}
\end{equation} 
with $\alpha^{\left(i\right)}_c$ being the value update step size at the $i$-th iteration. Notice that, $R\left(\boldsymbol{a}\left(t\right), \boldsymbol{\sigma}\left(t\right) \right) + \gamma \tilde V_{{\psi}^{\left(i\right)}}\left(\boldsymbol{\sigma}\left(t+1\right)\right)$ is a supervised term whose partial derivative will not be counted in the gradient update step defined in (\ref{eq:ACvupdate})\cite{sutton2018reinforcement}. 
 
 At the same time, based on the the policy gradient theorem \cite{sutton2000policy}, the ACRL algorithm updates policy function $\pi_{{\theta}^{\left(i\right)}}\left(\boldsymbol{a}\left(t\right)\left| \boldsymbol{\sigma}\left(t\right) \right.\right)$ by updating ${\theta}^{\left(i\right)}$ in the direction of the gradient of expected advantage as defined in
  \begin{equation}\label{eq:ACpupdate}
\begin{split}
{\theta}^{\left(i\right)}&={\theta}^{\left(i-1\right)}\\
&\;\;\;+\!\alpha^{\left(i-1\right)}_a\!\sum\limits_{t=1}^{T} {A}\left(\boldsymbol{a}\!\left(t\right)\!, \boldsymbol{\sigma}\!\left(t\right)\right)\nabla_{{\theta}}\log \pi_{{\theta}^{\left(i-1\right)}}\!\left(\boldsymbol{a}\!\left(t\right)\left| \boldsymbol{\sigma}\!\left(t\right)\right.\right),
\end{split}
\end{equation}
 where $\alpha^{\left(i\right)}_a$ is the policy update step size, at the $i$-th iteration of ACRL update. Note that, the ACRL algorithm updates policy and value functions with a mini-batch training procedure, i.e., it implements update after collecting a whole action-state-reward trajectory to reduce the variance on the system's learning performance caused by the action sampling, and the need on storing a big dataset in the system. The ACRL will repeat this trial, error and update procedure until a convergence to the optimal policy ${\pi}_{{\theta}^*}$ is reached. However, as we noticed, solving the high dimensional Dec-MDP in such a centralized way has high computational complexity. In the next subsection, we explain how our proposed D$^2$-RMRL algorithm solves the Dec-MDP with distributed routing policies ${\pi}_{{\theta}_f}$ on each file $f\in\mathcal{F}$, using the value decomposition technique. 

\subsection{Value Decomposition for Computing the Local Policies $\left\{\pi^f_0\left({a}^f\left(t\right)\left| {\sigma}^f\left(t\right) \right.\right), f=1,\ldots, N_{\textrm{F}}\right\}$ for a Given User Request Realization $\boldsymbol X$}

Our goal is to obtain optimal local policies  ${\pi}_{{\theta}_{f}}$ at each file $f$, which is a deep neural network parametrized by ${\theta}_{f}$. Each policy function ${\pi}_{{\theta}_{f}}$ takes file $f$'s local state ${\sigma}^f\left(t\right)\in \Omega^f$ as input, and outputs the probability of action ${a}^f\left(t\right)\in {\mathcal A}^f$ at current state.  To update these distributed policies locally for each file $f$, we decompose the original value function $V_{{\psi }}\left(\boldsymbol{\sigma}\left(t\right)\right)$ as the sum of local value functions, given by 
 \begin{equation}\label{eq:vd}
\begin{split}
V_{{\psi }}\left(\boldsymbol{\sigma}\left(t\right)\right) = \sum_{f\in\mathcal{F}}\tilde{V}_{{\psi}_{f}}\left({\sigma}^f\left(t\right)\right),
\end{split}
\end{equation} 
where $\tilde{V}_{{\psi}_{f}}\left({\sigma}^f\left(t\right)\right)$, parametrized by ${\psi}_{f}$, is the local value function associated with file $f$. 
Moreover, ${\eta}_f=\left\{{a}^f\left(t\right), {\sigma}^f\left(t\right), R^f\left({a}^f\left(t\right), {\sigma}^f\left(t\right) \right),t=1,\ldots, T\right\}$ is the action-state-reward trajectory generated by the single file routing process at file $f$ under current policy ${\pi}_{{\theta}_{f}}$. 
 Then the update of each value function is given by
{ \begin{equation}\label{eq:vupdate}
\setlength{\abovedisplayskip}{1 pt}
\setlength{\belowdisplayskip}{1 pt}
\begin{split}
{\psi}^{\left(i\right)}_{f}&={\psi}^{\left(i-1\right)}_{f}-\alpha^{\left(i\right)}_c\nabla_{{\psi}_{f}}\sum\limits_{t=1}^{T}A^2\left(\boldsymbol{a}\left(t\right), \boldsymbol{\sigma}\left(t\right)\right) \\
&={\psi}^{\left(i-1\right)}_{f}\!+\!2\alpha^{\left(i\right)}_c \!  \sum\limits_{t=1}^{T} \!A\left(\boldsymbol{a}\left(t\right), \boldsymbol{\sigma}\left(t\right)\right)\nabla_{{\psi}_{f}}\tilde{V}_{{\psi}^{\left(i-1\right)}_{f}}\left({\sigma}^f\left(t\right)\right). 
\end{split}
\end{equation}} 
Thus, using the value decomposition assumption in ({\ref{eq:vd}}), the gradient of each individual value function consists of a local term that depends on the local state ${\sigma}^f\left(t\right)$ and a  global term ${A}\left(\boldsymbol{a}\left(t\right), \boldsymbol{\sigma}\left(t\right)\right)$ that depends on the global actions $\boldsymbol{a}\left(t\right)$ and states $\boldsymbol{\sigma}\left(t\right)$. 
It is through this global term that the updates of all local value functions are coupled. Also, note that, $R\left(\boldsymbol{a}\left(t\right), \boldsymbol{\sigma}\left(t\right) \right) + \gamma \sum_{f\in\mathcal{F}} \tilde V_{{\psi}^{\left(i-1\right)}_{f}}\left({\sigma}^f\left(t+1\right)\right)$ is a supervised term whose partial derivative will not be counted in the gradient update step defined in (\ref{eq:vupdate}). 
Moreover, based on the the policy gradient theorem \cite{sutton2000policy}, the update on file $f$'s policy function parameters is given by
 \begin{equation}\label{eq:pupdate}
\begin{split}
{\theta}^{\left(i\right)}_{f}&={\theta}_{f}^{\left(i-1\right)}\\
&\;\;+\!\alpha^{\left(i-1\right)}_a\!\!\sum\limits_{t=1}^{T} \!{A}\left(\boldsymbol{a}\!\left(t\right)\!, \boldsymbol{\sigma}\!\left(t\right)\right)\nabla_{{\theta}_{f}}\!\log \pi_{{\theta}^{\left(i-1\right)}_{f}}\left({a}^f\!\!\left(t\right)\left| {\sigma}^f\!\!\left(t\right)\right.\right).
\end{split}
\end{equation}
Again we see that the updates of these local policies are coupled due to the global terms ${A}\left(\boldsymbol{a}\left(t\right), \boldsymbol{\sigma}\left(t\right)\right)$. 

{\color{black}  \begin{algorithm}[t]\small

\caption{Value decomposition-based ACRL algorithm for computing the local policies for a given user request realization $\boldsymbol X$.}   

\label{alg:D$^2$-RMRL}   
\setlength{\abovecaptionskip}{-40pt} 
\setlength{\belowcaptionskip}{-40pt}
\begin{algorithmic} [1] 
\REQUIRE User service requests $\boldsymbol x_u\left(t\right)$, $t=1, \ldots, T$, $u=1, \ldots, N_{\textrm{U}}$. \\ 
\vspace{2pt}  
\ENSURE Initialize value functions $\tilde V_{{\psi}^{\left(0\right)}_{f}}$, and policy functions ${\pi}_{{\theta}^{\left(0\right)}_{f}}$, for $f =1,\ldots,N_{\textrm{F}}$.\\

\vspace{2pt}  
\FOR {D$^2$-RMRL training epoch $i=1:I$} 
\vspace{2pt}  
\STATE Generate sample trajectories of state-action-reward ${\eta}_f\!=\!\left\{{\sigma}^f\left(t\right)\!, {a}^f\left(t\right)\!,  R^f\left({a}^f\left(t\right)\!, {\sigma}^f\left(t\right)\right)\!, t=1,\ldots,T\right\}$ for each $f =1,\ldots,N_{\textrm{F}}$ using the local policies $ \pi_{{\theta}^{\left(i-1\right)}_{f}}\left({a}^f\left(t\right)\left| {\sigma}^f\left(t\right)\right.\right)$ and Algorithm 1. 
\vspace{2pt}  
\STATE Calculate TD errors $A\left(\boldsymbol{a}\left(t\right), \boldsymbol{\sigma}\left(t\right) \right)$, $t=1,\ldots, T$ in (\ref{eq:tderror}).
\vspace{2pt}  
\FOR {each file $f=1:N_{\textrm{F}}$} 
\vspace{2pt}  
\STATE Update the local value function parameters ${\psi}^{\left(i\right)}_{f}$ using (\ref{eq:vupdate}).
\vspace{2pt}  
\STATE Update the local policy function parameters ${\theta}^{\left(i\right)}_{f}$ using (\ref{eq:pupdate}). 
\vspace{0.1pt}  
\ENDFOR
\vspace{2pt}  
\ENDFOR
\vspace{2pt}  
\RETURN{ Optimal value and policy functions, i.e., ${\psi}^{*}_{f}$, and ${\theta}^{*}_{f}$. }
\end{algorithmic}
\end{algorithm}}

In the considered system, the distributed policy and value functions are stored and optimized on a central controller. The resulting routing policy of file $f$ will be distributed to gateway $g$, if the routing process of file $f$ starts at gateway $g$. As summarized in Algorithm \ref{alg:D$^2$-RMRL}, starting with initial policy functions ${\pi}_{{\theta}^{\left(0\right)}_{f}}$ and value functions $\tilde V_{{\psi}^{\left(0\right)}_{f}}$, the algorithm uses the achieved overall pre-store hits (i.e., achieved reward) and the estimated reward (i.e., values of $\tilde{V}_{{\psi}_{f}}\left({\sigma}^f\left(t\right)\right)$) to calculate the TD error $A\left(\boldsymbol{a}\left(t\right), \boldsymbol{\sigma}\left(t\right)\right)$.  It then uses a mini-batch training mechanism to update policies and reward estimations (value functions) on each file $f$, independently, with (\ref{eq:vupdate}) and (\ref{eq:pupdate}), based a sample trajectory ${\eta}_f$. 
Such trial, error, then update procedure will be repeated, until the convergence is reached. 
Recall that the training process needs to be performed for each realization of the user request trajectory $\boldsymbol X$. To make the training process more efficient, we next resort to the technique of meta learning, to obtain a good initial network parameters $\left\{{\theta_f^{(0)}}, {\psi_f^{(0)}}, f= 1,\ldots, N_{\textrm{F}}\right\}$, for all $\boldsymbol X$ that follow a certain distribution $p(\boldsymbol X)$, such that starting from such meta-trained iniital parameters, Algorithm 1 will converge in just a few gradient iterations for any user request realization $\boldsymbol X \sim p(\boldsymbol X)$. 

 \subsection{Meta Training of Model Initials for a Given User Request Distribution $\boldsymbol X \sim p(\boldsymbol X)$}

%
In the previous subsection, we gave the ACRL algorithm for a given realization of the user request $\boldsymbol X$. In reality, $\boldsymbol X$ follows certain statistical distribution $p(\boldsymbol X)$. It is certainly not feasible to compute the policy for each possible realization of $\boldsymbol X$. To that end, we resort to meta learning. The basic idea is to train the networks' initial parameters $\left(\bar\theta_f, \bar\psi_f\right)$, $f=1, \ldots, N_{\textrm{F}}$, for the given distribution $p(\boldsymbol X)$, such that for any user request realizations $\boldsymbol X \sim p\left(\boldsymbol X\right)$, the corresponding optimal networks can be obtained from $\left(\bar\theta_f, \bar\psi_f\right)$, $f=1, \ldots, N_{\textrm{F}}$, through a few gradient update steps with a small amount of training data. 
In particular, the meta training procedure finds the learning initials, i.e., initial policy functions ${\pi}_{\bar{\theta}_f}$ and value functions $\tilde V_{\bar{\psi}_{f}}$, $f=1, \ldots, N_{\textrm{F}}$, that are already close to the optimal strategies and value estimations of all user request realizations following the distribution $p\left(\boldsymbol X\right)$. Starting from such learning initials, given a user request realization $\boldsymbol X$, the corresponding optimal policy and value functions can be obtained by a few actor-critic updates. 

At each meta training update epoch, we use $J$ samples of user request $\boldsymbol X$, i.e., $\boldsymbol X_1, \ldots, \boldsymbol X_J \sim  p\left(\boldsymbol X\right)$, to update $\bar{\psi}_{f}$, and ${\bar{\theta}_{f}}$ in a similar way as in (\ref{eq:vupdate}) and (\ref{eq:pupdate}), except that now $J$ state-action-reward trajectories are used, one for each $\boldsymbol X_j$. In particular, we can write, for $f=1, \ldots, N_{\textrm{F}}$,
{ \begin{equation}\label{eq:vupdatem}
 \setlength{\abovedisplayskip}{3 pt}
\setlength{\belowdisplayskip}{-1 pt}
\begin{split}
\bar{\psi}_{f}\leftarrow\bar{\psi}_{f}+2\alpha_c  \!\sum\limits_{j=1}^{J} \sum\limits_{t=1}^{T} A\left(\boldsymbol{a}_j\left(t\right)\!, \boldsymbol{\sigma}_j\left(t\right)\right)\nabla_{\bar{\psi}_{f}}\tilde{V}_{\bar{\psi}_{f,j}}\left({\sigma}^f_j\left(t\right)\right),
\end{split}
\end{equation}} 
 \begin{equation}\label{eq:pupdatem}
  \setlength{\abovedisplayskip}{3 pt}
\setlength{\belowdisplayskip}{1 pt}
\begin{split}
{\bar{\theta}_{f}} &\leftarrow {\bar{\theta}_{f}}\\
 &\;\;+\alpha_a\!\sum\limits_{j=1}^{J}\!\sum\limits_{t=1}^{T} \!{A}\left(\boldsymbol{a}_j\!\left(t\right), \boldsymbol{\sigma}_j\!\left(t\right)\right)\nabla_{{\bar{\theta}_{f}}}\!\log \pi_{{\bar{\theta}_{f,j}}}\!\left({a}^f_j\!\left(t\right)\left| {\sigma}^f_j\!\left(t\right)\right.\right),
\end{split}
\end{equation}
where $\bar{\psi}_{f,j}$, and ${\bar{\theta}_{f,j}}$ are, respectively, the value and policy function parameters updated at each user request $\boldsymbol X_j$. ${\sigma}^f_j\left(t\right)$, and ${ a}^f_j\left(t\right)$, are the state and action sampled at time slot $t$ corresponding to user request $\boldsymbol X_j$, respectively.  To obtain the network parameters $\bar{\psi}_{f,j}$, ${\bar{\theta}_{f,j}}$, and the state-action-reward trajectory ${ \eta}_{f,j}=\left\{{\sigma}^f_j\left(t\right), { a}^f_j\left(t\right),  R^f\left({ a}^f_j\left(t\right),  {\sigma}^f_j\left(t\right)\right), t=1,\ldots, T\right\}$ for each $\boldsymbol X_j$, we proceed as follow. First, we sample the state-action-reward trajectory using the current policy function $\pi_{{\bar{\theta}_{f}}}$ and user request $\boldsymbol X_j$ to obtain ${\bar \eta}_{f,j}=\left\{{\bar\sigma}^f_j\left(t\right), {\bar a}^f_j\left(t\right),  R^f\left({\bar a}^f_j\left(t\right), \bar {\sigma}^f_j\left(t\right)\right), t=1,\ldots, T\right\}$.
Then we update the model parameters using one-step gradient descent as, for , $f=1, \ldots, N_{\textrm{F}}$ 
{ \begin{equation}\label{eq:vupdatem1}
 \setlength{\abovedisplayskip}{3 pt}
\setlength{\belowdisplayskip}{-1 pt}
\begin{split}
\bar{\psi}_{f,j}=\bar{\psi}_{f}+2\alpha_c   \sum\limits_{t=1}^{T} A\left(\boldsymbol{\bar a}_j\left(t\right), \boldsymbol{\bar \sigma}_j\left(t\right)\right)\nabla_{\bar{\psi}_{f}}\tilde{V}_{\bar{\psi}_{f}}\left(\bar{\sigma}^f_j\left(t\right)\right),
\end{split}
\end{equation}} 
 \begin{equation}\label{eq:pupdatem1}
  \setlength{\abovedisplayskip}{3 pt}
\setlength{\belowdisplayskip}{1 pt}
\begin{split}
{\bar{\theta}_{f,j}}&= {\bar{\theta}_{f}}\\
 &\;\;+\alpha_a\sum\limits_{t=1}^{T} {A}\left(\boldsymbol{\bar a}_j\left(t\right), \boldsymbol{\bar \sigma}_j\left(t\right)\right)\nabla_{{\bar{\theta}_{f}}}\log \pi_{{\bar{\theta}_{f}}}\left(\bar{a}^f_j\left(t\right)\left| \bar{\sigma}^f_j\left(t\right)\right.\right).
\end{split}
\end{equation}

Next, we generate sample trajectory ${ \eta}_{f,j}$ using the updated policy $\pi_{{\bar{\theta}_{f,j}}}$, so as to update learning initials with (\ref{eq:vupdatem}), and (\ref{eq:pupdatem}). 
As summarized in Algorithm \ref{alg: meta}, at each meta training iteration, $J$ user request realizations are sampled from $p\left(\boldsymbol{X}\right)$. Using each realization, $\boldsymbol{X}_j\sim p\left(\boldsymbol{X}\right)$, for each file $f$, we obtain the sample trajectory $\bar\eta_{f,j}$ using the current policy $\pi_{\bar{\theta}_{f}}$, and then perform one-step update on the value and policy functions using (\ref{eq:vupdatem1}) and (\ref{eq:pupdatem1}). Next, using the updated policy $\pi_{\bar{\theta}_{f,j}}$, we obtain the sample trajectory ${\eta}_{f,j}$. Finally, the initial policy function $\pi_{\bar \theta_f}$ and value function $V_{\bar \psi_f}$ are updated based on (\ref{eq:pupdatem}) and (\ref{eq:vupdatem}), respectively. 

 In essence, the above meta training procedure seeks to find the optimal learning initializations, i.e., ${\pi}_{\bar{\theta}^{\left(0\right)}_{f}}={\pi}_{\bar{\theta}_{f}}$, and $V_{\bar{\psi}^{\left(0\right)}_{f}}=V_{\bar{\psi}_{f}}$, that are close to the optimal policies and values for all user request realizations. Starting from these initializations, the proposed D$^2$-RMRL solution, i.e., Algorithm 2, takes only a few iterations to reach convergence for every possible user requests $\boldsymbol{X} \sim p\left(\boldsymbol{X}\right)$. 
Given the meta trained policies $\pi_{\bar\theta_f}$ and value functions  $V_{\bar\psi_f}$, in order to obtain the optimal policy for any given $\boldsymbol X \sim p(\boldsymbol X)$, we simply run Algorithm 2 by initializing ${\pi}_{\bar{\theta}^{\left(0\right)}_{f}}={\pi}_{\bar{\theta}_{f}}$, and $V_{\bar{\psi}^{\left(0\right)}_{f}}=V_{\bar{\psi}_{f}}$, $f=1, \ldots, N_{\textrm{F}}$. Then the number of training epochs needed for reaching convergence is typically small.

\begin{algorithm}[t]\small
\caption{Meta training for optimal learning initials for a given user request distribution $p\left(\boldsymbol{X}\right)$.}
\label{alg: meta}   
\setlength{\abovecaptionskip}{-40pt} 
\setlength{\belowcaptionskip}{-40pt}
\begin{algorithmic} [1] 
\REQUIRE User request distribution $p\left(\boldsymbol{X}\right)$. \\ 
\vspace{2pt}  
\ENSURE Initialize value functions $\tilde{V}_{\boldsymbol{\bar \psi}_{f}}$, and policy functions ${\pi}_{\boldsymbol{\bar \theta}_{f}}$, for $f =1,\ldots,N_{\textrm{F}}$.\\ 
\vspace{2pt}  
\FOR {Meta training epoch $i=1:I$} 
\vspace{2pt}  
\FOR {$j=1:J$} 
\vspace{2pt}  
\STATE Sample user request realization $\boldsymbol{X}_j\sim p\left(\boldsymbol{X}\right)$.
\vspace{2pt}  
\STATE Generate sample trajectories of state-action-reward ${\bar \eta}^f_j\!=\!\left\{{\bar \sigma}^f_j\left(t\right)\!, {\bar a}^f_j\left(t\right)\!,  R^f\left({\bar a}^f_j\left(t\right)\!, {\bar \sigma}^f_j\left(t\right)\right), t=1,\ldots,T\right\}$ using the initial policy functions ${\pi}_{{\bar \theta}_{f}}$ and Algorithm 1, for $f =1,\ldots,N_{\textrm{F}}$. 
\vspace{2pt}  
\STATE Calculate $A\left(\boldsymbol{\bar a}\left(t\right), \boldsymbol{\bar \sigma}\left(t\right) \right)$ in (\ref{eq:vupdatem1}) and (\ref{eq:pupdatem1}), $t=1,\ldots, T$ using (\ref{eq:tderror}).
\vspace{2pt}  
\FOR {Each file $f=1:N_{\textrm{F}}$} 
\vspace{2pt}  
\STATE Perform one-step update on the value and policy functions using (\ref{eq:vupdatem1}) and (\ref{eq:pupdatem1}), to obtain $\bar{\psi}_{f,j}$, and $\bar{\theta}_{f,j}$.
\vspace{2pt}  
\STATE Generate the state-action-reward trajectory ${\eta}^f_j\!=\!\left\{{\sigma}^f_j\left(t\right)\!, {a}^f_j\left(t\right)\!,  R^f\left({a}^f_j\left(t\right)\!, {\sigma}^f_j\left(t\right)\right), t=1,\ldots,T\right\}$ using the updated policy functions ${\pi}_{\bar{\theta}_{f,j}}$ and Algorithm 1. 
\vspace{2pt}  
\ENDFOR
\vspace{2pt}  
\STATE Calculate $A\left(\boldsymbol{a}_j\left(t\right), \boldsymbol{\sigma}_j\left(t\right) \right)$ in (\ref{eq:vupdatem}) and (\ref{eq:pupdatem}), $t=1,\ldots, T$ using (\ref{eq:tderror}).
\vspace{2pt}  
\ENDFOR
\vspace{2pt} 
\FOR {Each file $f=1:N_{\textrm{F}}$} 
\vspace{2pt}  
\STATE Update initial value parameters $\bar{\psi}_{f}$ using (\ref{eq:vupdatem}), 
and policy parameters $\bar{\theta}_{f}$ using (\ref{eq:pupdatem}). 
\ENDFOR
\vspace{2pt}  
\ENDFOR
\vspace{2pt}  
\RETURN{ Optimal initial policy functions ${\pi}_{{\theta}^{\left(0\right)}_{f}}={\pi}_{{\bar \theta}_{f}}$, and initial value functions $\tilde{V}_{{\psi}^{\left(0\right)}_{f}}=\tilde{V}_{{\bar \psi}_{f}}$, for $f =1,\ldots,N_{\textrm{F}}$}. 
\end{algorithmic}
\end{algorithm} 

 \subsection{Pre-training for Distribution Robust Meta Learning}
In practice, the user request distribution may vary, e.g., it depends on different times of the day. Assume that there are totally $K$ user request distributions $p_1\left(\boldsymbol X\right), \ldots, p_K\left(\boldsymbol X\right)$. Then we can simply apply Algorithm \ref{alg: meta} to perform meta training for each one of these distributions to obtain the corresponding initials. However, such independent meta training is time-consuming and we would like to make use of the initials already meta trained for some distributions, to speed up the entire meta training process for all distributions. 
Specifically, suppose meta training is sequentially performed for $p_1\left(\boldsymbol X\right), ..., p_K\left(\boldsymbol X\right)$. At the beginning of meta training for $p_k(\boldsymbol X)$, we already have the meta trained initials $\left(\bar\psi^\ell, \bar\theta^\ell\right)$ for $\ell =1, \ldots, k-1$, 
where $\bar\psi^\ell = \left[ \bar\psi_f^\ell, f=1, \ldots, N_F\right]$, and $\bar\theta^\ell = \left[ \bar\theta_f^\ell, f=1, \ldots, N_F\right] $.
Then, when we perform meta training for $p_k\left(\boldsymbol X\right)$ using Algorithm \ref{alg: meta}, instead of randomly initializing $\bar\psi$ and $\bar\theta$, we start with one of the previous meta trained initials, based on the ``distance'' between each initial to the optimum under $p_k \left(\boldsymbol X\right)$. 

In particular,  we obtain $J$ samples of user request realizations $\boldsymbol X_j \sim p_k\left(\boldsymbol X\right), j=1, \ldots, J$. For each of these realizations $\boldsymbol X_j$, we apply each available initial policy functions $\pi_{{\bar{\theta}^\ell_{f}}}$ to obtain the state-action-reward trajectories ${\bar \eta}^\ell_{f,j}=\left\{{\bar\sigma}^\ell_{f,j}\left(t\right), {\bar a}^\ell_{f,j}\left(t\right),  R\left({\bar a}^\ell_{f,j}\left(t\right), \bar {\sigma}^\ell_{f,j}\left(t\right)\right), t=1,\ldots, T\right\}$, $f=1,\ldots, N_{\textrm{F}}$. 
Then, we update the model parameters using one-step gradient descent as 
{ \begin{equation}\label{eq:vupdatem2}
\setlength{\abovedisplayskip}{3 pt}
\setlength{\belowdisplayskip}{-1 pt}
\begin{split}
\bar{\psi}^\ell_{f,j}\leftarrow\bar{\psi}^\ell_{f}+2\alpha_c   \sum\limits_{t=1}^{T} A\left(\boldsymbol{\bar a}^\ell_j\left(t\right), \boldsymbol{\bar \sigma}^\ell_j\left(t\right)\right)\nabla_{\bar{\psi}^\ell_{f}}\tilde{V}_{\bar{\psi}^\ell_{f}}\left(\bar{\sigma}^\ell_{f,j}\left(t\right)\right),
\end{split}
\end{equation}} 
 \begin{equation}\label{eq:pupdatem2}
 \setlength{\abovedisplayskip}{3 pt}
\setlength{\belowdisplayskip}{3 pt}
\begin{split}
{\bar{\theta}^\ell_{f,j}} &\leftarrow {\bar{\theta}^\ell_{f}}\\
& +\alpha_a\sum\limits_{t=1}^{T}\! {A}\left(\boldsymbol{\bar a}^\ell_j\left(t\right), \boldsymbol{\bar \sigma}^\ell_j\left(t\right)\right)\!\nabla_{{\bar{\theta}^\ell_{f}}}\log \pi_{{\bar{\theta}^\ell_{f}}}\left(\bar{a}^\ell_{f,j}\left(t\right)\left| \bar{\sigma}^\ell_{f,j}\left(t\right)\right.\right).
\end{split}
\end{equation} 
At the next step, we generate trajectory ${ \eta}^\ell_{f,j}=\left\{{\sigma}^\ell_{f,j}\left(t\right), { a}^\ell_{f,j}\left(t\right),  R\left({ a}^\ell_{f,j}\left(t\right),  {\sigma}^\ell_{f,j}\left(t\right)\right), t=1,\ldots, T\right\}$, using the updated policy $\pi_{{\bar{\theta}^\ell_{f,j}}}$, so as to calculate the ``distance'' between each initial $\left(\bar\psi^\ell, \bar\theta^\ell\right)$ to the optimum under $p_k \left(\boldsymbol X\right)$ as in
 \begin{equation}\label{eq:plossP}
\begin{split}
d_{\theta,f}^\ell  \buildrel \Delta \over = & \sum^{J}_{j=1}{{{ A}}} \left( \boldsymbol{a}^\ell_j\left(t\right),\boldsymbol{\sigma}^\ell_j\left(t\right)\right)   \pi_{\bar{\theta}_{f,j}}\left( a^\ell_{f,j}\left(t\right)\left| {\sigma}^\ell_{f,j}\left(t\right) \right.\right),
\end{split}
\end{equation}
\begin{equation}\label{eq:clossP}
\begin{split}
{d_{\psi,f}^\ell \buildrel \Delta \over = \sum^{J}_{j=1}\left(R\left({a}^\ell_{f,j}\left(t\right)\left|{\sigma}^\ell_{f,j}\left(t\right)\right.\right)- \tilde{V}_{\bar{\psi}_{f,j}}\left({\sigma}^\ell_{f,j}\left(t\right)\right) \right)^2}.
\end{split}
\end{equation}
The distance between the $\ell$-th initial and the optimum for $p_k\left(\boldsymbol{X}\right)$ is then given by $D_k\left(\ell\right) = \sum\limits_{f=1}^{N_F} \left(d_{\theta,f}^\ell+ d_{\psi,f}^\ell\right)$, $\ell = 1, ..., k-1$. Then, the meta training procedure for service distribution $p_k\left(\boldsymbol{X}\right)$ is initialized by $\left(\bar \psi^{\ell^*}, \bar\theta^{\ell^*}\right)$, where $\ell^* = \arg\min_{\ell} D_k\left(\ell\right)$.

{\color{black}  \begin{algorithm}[t]\small

\caption{Pre-training for shortened meta training for different user request distributions.}   

\label{alg:pre-train}   
\setlength{\abovecaptionskip}{-40pt} 
\setlength{\belowcaptionskip}{-40pt}
\begin{algorithmic} [1] 
\REQUIRE User request distributions $p_1\left(\boldsymbol{X}\right), \ldots, p_K\left(\boldsymbol{X}\right)$. \\ 
\vspace{2pt}  
\ENSURE $k=1$: Run Algorithm \ref{alg: meta} to obtain the meta trained initial $\left(\bar\psi^1, \bar\theta^1\right)$ for $p_1\left(\boldsymbol{X}\right)$. \\

\vspace{2pt}  
\vspace{2pt}  
\FOR {$k=2:K$} 
\FOR {environment sampling epoch $j=1:J$} 
\vspace{2pt}  
\STATE Sample a user request realization $\boldsymbol{X}_j\sim P_{k}\left(\boldsymbol{X}\right)$.
\vspace{2pt}  
\FOR {treated user request distributions $\ell =1:k-1$} 
\vspace{2pt}  
\STATE Generate sample trajectories of state-action-reward ${\bar \eta}^\ell_{f,j}=\left\{{\bar\sigma}^\ell_{f,j}\left(t\right), {\bar a}^\ell_{f,j}\left(t\right),  R\left({\bar a}^\ell_{f,j}\left(t\right), \bar {\sigma}^\ell_{f,j}\left(t\right)\right), t=1,\ldots, T\right\}$ using the initial policy functions ${\pi}_{{\bar \theta}^\ell_{f}}$ and Algorithm 1, for $f =1,\ldots,N_{\textrm{F}}$. 
\vspace{2pt}  
\STATE Calculate $A\left(\boldsymbol{\bar a}^\ell_j\left(t\right), \boldsymbol{\bar \sigma}^\ell_j\left(t\right) \right)$ in (\ref{eq:vupdatem2}) and (\ref{eq:pupdatem2}), $t=1,\ldots, T$, using (\ref{eq:tderror}).
\vspace{2pt}  
\FOR {Each file $f=1:N_{\textrm{F}}$} 
\vspace{2pt}  
\STATE Perform one-step update on the value and policy functions using (\ref{eq:vupdatem2}) and (\ref{eq:pupdatem2}), to obtain $\bar{\psi}^\ell_{f,j}$, and $\bar{\theta}^\ell_{f,j}$.
\vspace{2pt}  
\STATE Generate the state-action-reward trajectory ${ \eta}^\ell_{f,j}=\left\{{\sigma}^\ell_{f,j}\!\left(t\right), { a}^\ell_{f,j}\!\left(t\right)\!,  R\left({ a}^\ell_{f,j}\!\left(t\right),  {\sigma}^\ell_{f,j}\!\left(t\right)\right)\!, t=1,\ldots, T\right\}$ using the updated policy functions ${\pi}_{\bar{\theta}^\ell_{f,j}}$ and Algorithm 1. 
\vspace{2pt}  
\ENDFOR
\vspace{2pt}  
\STATE Calculate $A\left(\boldsymbol{a}^\ell_j\left(t\right), \boldsymbol{\sigma}^\ell_j\left(t\right) \right)$ in (\ref{eq:plossP}) and (\ref{eq:clossP}), $t=1,\ldots, T$ using (\ref{eq:tderror}).
\vspace{2pt}  
\ENDFOR

\vspace{2pt}  
\ENDFOR
\vspace{2pt}  
\STATE Compute (\ref{eq:plossP}), (\ref{eq:clossP}), and then $D_k\left(\ell\right)$, for all $\ell =1,\ldots, k-1$. 
\vspace{2pt}  
\STATE Compute $\ell^*$.  Run Algorithm \ref{alg: meta} using the initializations $\left(\bar \psi^{\ell^*}, \bar\theta^{\ell^*}\right)$ to obtain the meta trained initials $\left({\pi}_{{\bar \theta}^k_{f}},  \tilde{V}_{{\bar \psi}^k_{f}}\right)$, for $f =1,\ldots,N_{\textrm{F}}$, and $p_k\left(\boldsymbol{X}\right)$. 
\vspace{2pt}  
\ENDFOR
\vspace{2pt}  
\RETURN{ Optimal meta training initials $\left(\bar \psi^{k}, \bar\theta^{k}\right)$, for $k=1,\ldots,K$ }. 
\end{algorithmic}
\end{algorithm} }
The proposed D$^2$-RMRL with pre-training is summarized in Algorithm \ref{alg:pre-train}. For the first service request distribution $p_1(\boldsymbol {X})$, the algorithm randomly starts a meta training procedure, using Algorithm \ref{alg: meta}. Then for each subsequent new distributions, the algorithm can achieve a shortened meta training by choosing among the learning initials corresponding to distributions that are already meta trained. In particular, it collects experience on serving unseen user request distributions using the available learning initials from treated distributions, then obtains the update with  (\ref{eq:vupdatem2}) and (\ref{eq:pupdatem2}), and evaluates the update with distance metrics in (\ref{eq:plossP}) and (\ref{eq:clossP}). Finally, it will use the best one among the available learning initials to start the meta training procedure for the current distribution. Through such a transfer process of learned initials, the overall meta training process over multiple user request distributions can be expedited.

\section{Simulation Results} 
\subsection{Simulation Setup}
{
For our simulations, we consider a scenario with $N_{\textrm{G}}=5$ gateways serving $N_{\textrm{U}}=20$ user clusters with the help of a LEO cube satellite constellation at the altitude of $550$ km with an inclination of $53^{\circ}$. 
In particular, these user clusters and gateways fall into the service coverage of $N_{\textrm{S}}= 12$ satellites on $4$ intertwined orbits of the constellation. Based on the satellite orbit information in \cite{Tracker} and ground device locations, we construct a time-unrolled data transmission graph to capture the contact chances in the system,  within $T=100$ of $10$-second time slots. 
Within this graph, a user cluster or a ground gateway can only contact with its on-duty satellite, i.e., the satellite that is serving their corresponding active cell as in \cite{Update}. Meanwhile, we assume that two satellites can communicate only when the distance between them is less than one active cell diameter \cite{Tracker}.  
Moreover, we assume there are in total $N_{\textrm{F}}=15$ on-request files in the system. 
The user request $\boldsymbol{X} = \left\{ x_u^{f}\left(t\right), u=1,\ldots, N_{\textrm{U}}, f = 1, \ldots, N_{\textrm{F}}, t=1,\ldots, T\right\}$ are generated as follows. At each time slot, each user cluster $u$ generates $m_u$ file requests, where $m_u \in \left\{0, 1, \ldots, N_F\right\}$  follows a truncated Poisson distribution with mean ${\bar m}_u$; and these $m_u$ files are random selected out of the $N_F$ files for which  we set $x_u^{f}\left(t\right) =1$. Different user request distributions correspond to different mean values of $m$. 


The value and policy functions of the D$^2$-RMRL algorithm are both represented by feed forward neural networks, with $2$ hidden layers, each is composed of  $100$ elements. The results of proposed D$^2$-RMRL algorithm are compared with the ones of the independent actor-critic (IAC) algorithm \cite{foerster2017counterfactual}, 
randomly initialized value decomposition RL solution described in Section III.B (denoted as RL), and meta trained value decomposition RL described in Section III.C (denoted as MRL). 
Recall that the proposed VD-RL algorithm updates policy and value functions locally at each files based on the global term $A\left(\boldsymbol{a}\left(t\right), \boldsymbol{\sigma}\left(t\right) \right)$, as in (\ref{eq:vupdate}) and (\ref{eq:pupdate}). In contrast,  the IAC algorithm replaces this global term $A\left(\boldsymbol{a}\left(t\right), \boldsymbol{\sigma}\left(t\right) \right)$ in (\ref{eq:vupdate}) and (\ref{eq:pupdate}) with the local term given by $R^f\left({a}^f\left(t\right), {\sigma}^f\left(t\right) \right)+\gamma V_{{\psi }^{f}}\left({\sigma}^f\left(t+1\right)\right)- V_{{\psi }^{f}}\left({\sigma}^f\left(t\right)\right)$. 
Thus, the comparison between IAC and the proposed solution can justify how the proposed value decomposition solution improves distributed data transmission control in the considered satellite network. Meanwhile, the results of the proposed algorithm are also compared to the ones from the RL and MRL solutions, which demonstrate how the proposed distribution-robust meta training mechanism shortens the learning based data transmission design procedures. 
All statistical results are averaged over a large number of independent runs.

\subsection{Evaluation of Algorithm \ref{alg:D$^2$-RMRL} -- Value Decomposition} 
We first evaluate the performance of value decomposition RL solution in Algorithm \ref{alg:D$^2$-RMRL} for one user request realization $\boldsymbol{X}$. In Fig. \ref{VD}, we show the convergence behaviors of Algorithm \ref{alg:D$^2$-RMRL} and the IAC method, with the shades indicating results of $1000$ runs of the algorithms with random initializations for the same  user request realization $\boldsymbol{X}$. Fig. \ref{VD} shows that, on average, the value decomposition approach proposed in Section III.B yields a $31.8\%$ higher final pre-store hits than the IAC method, as it reinforces strategies that benefit the whole team.  On the other hand, the IAC method can only find strategies that maximize the individual utilities. Moreover, from Fig. \ref{VD}, we also see that Algorithm \ref{alg:D$^2$-RMRL} converges much faster than the IAC method. 
\begin{figure}[t]
  \begin{center}
   \vspace{0cm}
    \includegraphics[width=8.3 cm]{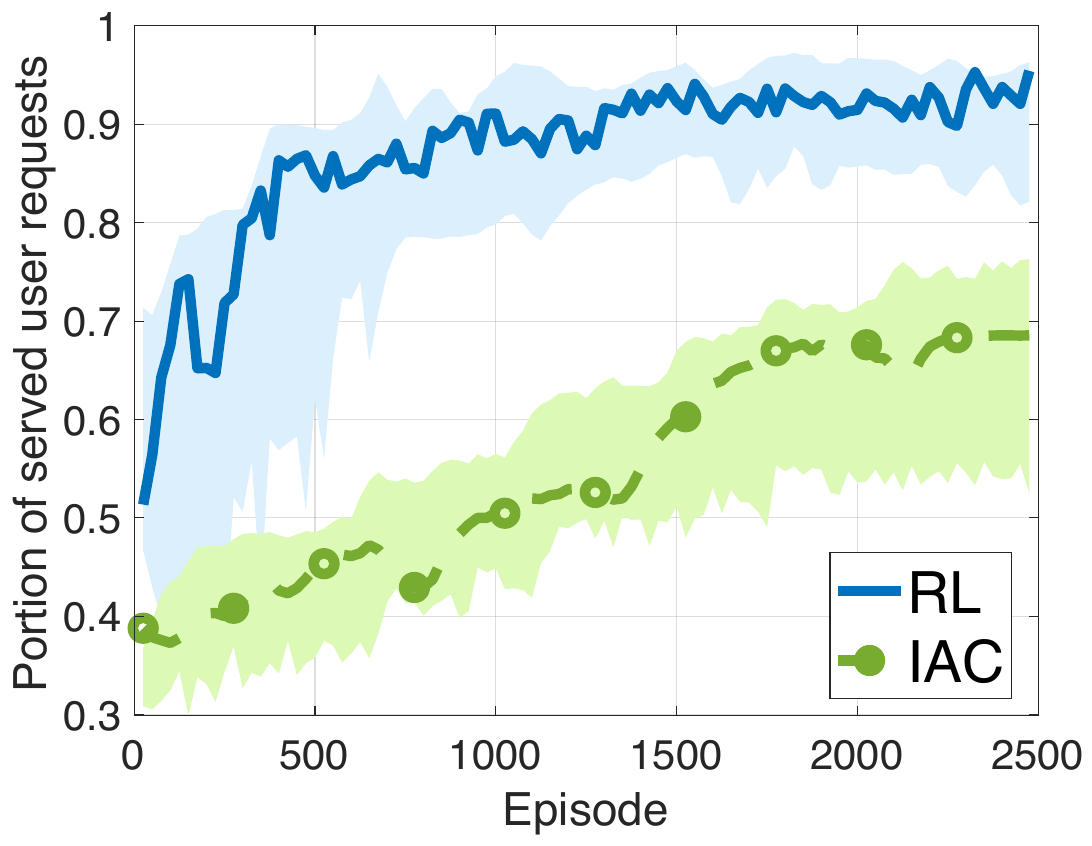}
    \caption{\label{VD} {Convergence comparison between Algorithm \ref{alg:D$^2$-RMRL} and the IAC method for a fixed user request realization.}}
  \end{center}
\end{figure}

\subsection{Evaluation of Algorithm \ref{alg: meta} -- Meta Training} 
\begin{figure}[t]
  \begin{center}
   \vspace{0cm}
    \includegraphics[width=8.3cm]{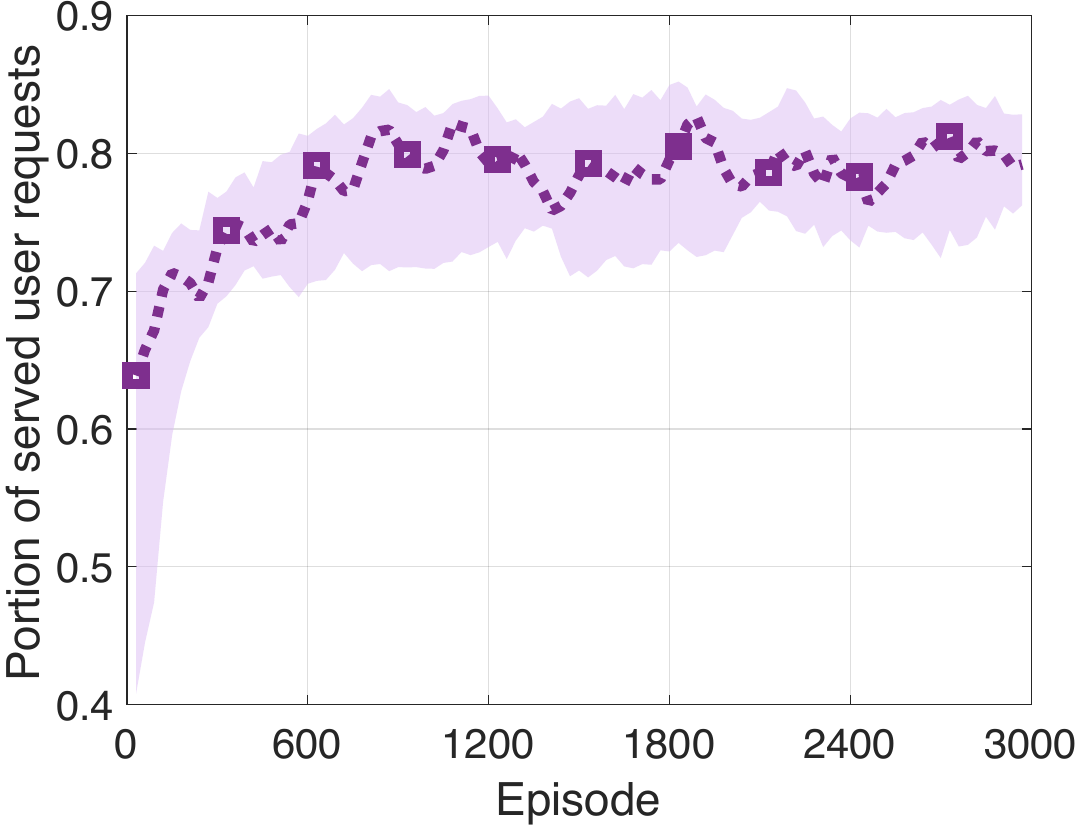}
    \caption{\label{metacvg} {Convergence of Algorithm \ref{alg: meta}.}}
  \end{center}
\end{figure}
\begin{figure}[t]
  \begin{center}
   \vspace{0cm}
    \includegraphics[width=8.3cm]{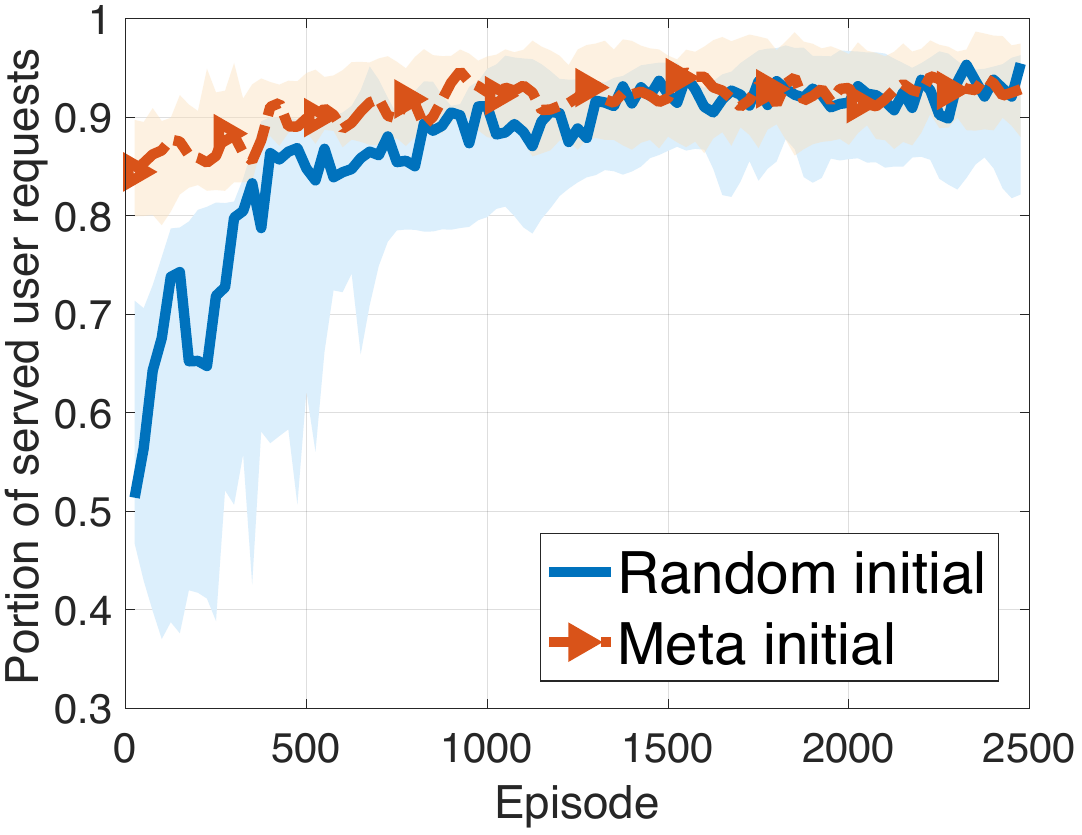}
    \caption{\label{meta} {Convergence of Algorithm \ref{alg:D$^2$-RMRL} under meta initial given by Algorithm \ref{alg: meta}, and random initial.}}
  \end{center}
\end{figure}
Next we evaluate how the meta training technique in Algorithm \ref{alg: meta} shortens the learning procedure of Algorithm \ref{alg:D$^2$-RMRL} for user requests realizations following $\boldsymbol{X} \sim P\left(\boldsymbol{X} \right)$, with ${\bar m}_u=1$ for $u=1,\cdots,10$, ${\bar m}_u=2$ for $u=11,\cdots,20$. Firstly, Figure \ref{metacvg} shows the convergence behavior of the meta training procedure, i.e., Algorithm \ref{alg: meta} and it is seen that converge can be reached in about $1100$ iterations on average. The shades in Figure \ref{metacvg} and Figure \ref{meta} indicate results of $1000$ independent runs with random sample user requests from distribution $P\left(\boldsymbol{X} \right)$. 
Figure \ref{meta} shows that, starting from the meta trained learning initials given by Algorithm \ref{alg: meta}, Algorithm \ref{alg:D$^2$-RMRL} takes about $920$ iterations to reach convergence, which improves the convergence speed by up to $40.7\%$ compared to random initials. This stems from the fact that, with the meta training initialization, the RL algorithm can start from policies that are close to the team optimal strategies for the target service task. Moreover, by comparing the shaded areas of the two curves, it is seen that using the meta initial can considerably reduce the variation of the performance of Algorithm \ref{alg:D$^2$-RMRL}, in addition to speeding up the convergence.

%
\subsection{Evaluation of Algorithm \ref{DRMETAC} -- Pre-training} 
In Fig. \ref{DRMETAC}, we show how Algorithm \ref{alg:pre-train} speeds up the meta training convergence for a family of distributions. We assume that there are $K=4$ service distributions $P_1\left(\boldsymbol{X} \right),\ldots, P_4\left(\boldsymbol{X} \right)$ and we need to obtain the meta initial for each $P_i\left(\boldsymbol{X} \right)$. In Fig. \ref{DRMETAC}, we compare the convergence behaviors of two approaches: one is Algorithm \ref{alg:pre-train} and the other is running Algorithm \ref{alg: meta} four times one for each $P_i\left(\boldsymbol{X} \right)$. Recall that in Algorithm \ref{alg:pre-train}, for the first distribution $P_1\left(\boldsymbol{X} \right)$, it simply runs Algorithm \ref{alg: meta}. Hence the convergence behavior  for $P_1\left(\boldsymbol{X} \right)$ is the same for both approaches. Then, for the other service distributions, $P_2\left(\boldsymbol{X} \right),P_3\left(\boldsymbol{X} \right), P_4\left(\boldsymbol{X} \right)$, by making use of the meta initial already obtained for the distribution that is closest to the current one, Algorithm \ref{alg:pre-train}, can converge up  $43.7\%$ faster than Algorithm \ref{alg:pre-train} which always starts from random initials.  

\begin{figure}[t]
  \begin{center}
   \vspace{0cm}
    \includegraphics[width=8.3 cm]{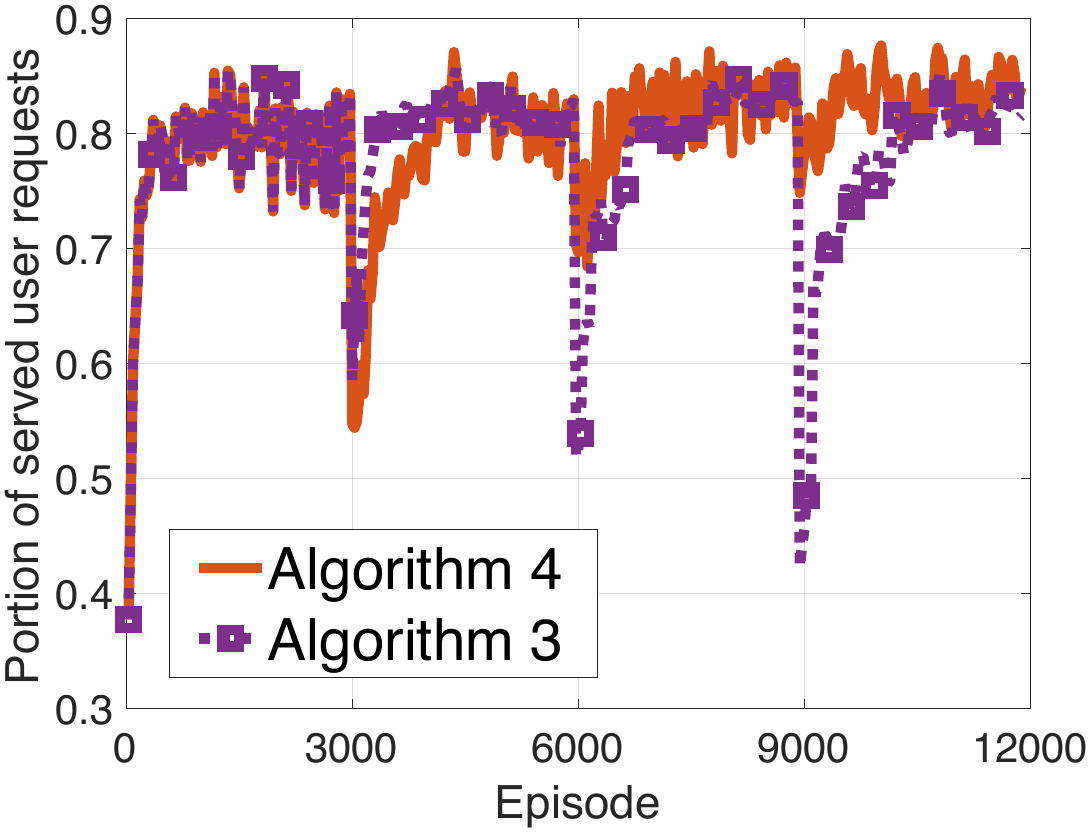}
    \caption{\label{DRMETAC} {Convergence of meta training for a family of service distributions by Algorithm \ref{DRMETAC}, and Algorithm \ref{alg: meta} with random initialization for each distribution.} }
  \end{center}
\end{figure}

}
\section{Conclusions}
{
In this paper, we have studied the problem of pre-storing and routing data to satellites in a cube satellite network. Using this network, the ground users' dynamic and unforeseeable data requests are served by the cube satellites, which pre-store data from distributed ground gateways, and deliver data service to users in its coverage areas. The design problem is to determine the data to be pre-stored in each satellite and how to route it from a gateway to the satellite.  
 We have formulated this problem as Dec-MDP and have proposed a D$^2$-RMRL algorithm to solve this problem. The proposed D$^2$-RMRL algorithm is based on a multi-agent reinforcement learning approach and makes use of the value decomposition technique, so that the agents  independently optimize their strategies toward the maximal overall pre-store hits, by sharing only their achieved and estimated reward with each other. 
To reduce the excessive training cost of this machine learning based solution for different user service requests, we have proposed  the meta trainging procedure to obtain initials that can significantly speed up the training process for a given service request distribution, as well as a pre-training procedure for further speedup the meta training procedure for a family of different service request distributions. 
Simulation results show that the proposed D$^2$-RMRL algorithm achieves high rate of pre-store hits with fast convergence. 
}

\bibliographystyle{IEEEbib}
\bibliography{references}

\begin{thebibliography}{10}

\bibitem{maral2020satellite}
G.~Maral, M.~Bousquet, and Z.~Sun,
\newblock {\em Satellite communications systems: systems, techniques and
  technology},
\newblock John Wiley \& Sons, 2020.

\bibitem{7956007}
M.~Sheng, Y.~Wang, J.~Li, Ru. Liu, D.~Zhou, and L.~He,
\newblock ``Toward a flexible and reconfigurable broadband satellite network:
  Resource management architecture and strategies,''
\newblock {\em IEEE Wireless Communications}, vol. 24, no. 4, pp. 127--133,
  Aug. 2017.

\bibitem{8906029}
J.~A. Fraire, G.~Nies, C.~Gerstacker, H.~Hermanns, K.~Bay, and M.~Bisgaard,
\newblock ``Battery-aware contact plan design for {LEO} satellite
  constellations: The {Ulloriaq} case study,''
\newblock {\em IEEE Transactions on Green Communications and Networking}, vol.
  4, no. 1, pp. 236--245, Mar. 2020.

\bibitem{7883913}
D.~Zhou, M.~Sheng, X.~Wang, C.~Xu, R.~Liu, and J.~Li,
\newblock ``Mission aware contact plan design in resource-limited small
  satellite networks,''
\newblock {\em IEEE Transactions on Communications}, vol. 65, no. 6, pp.
  2451--2466, Jun. 2017.

\bibitem{fraire2021scalability}
J.~A Fraire, C.~Gerstacker, H.~Hermanns, G.~Nies, M.~Bisgaard, and K.~Bay,
\newblock ``On the scalability of battery-aware contact plan design for {LEO}
  satellite constellations,''
\newblock {\em International Journal of Satellite Communications and
  Networking}, vol. 39, no. 2, pp. 193--204, Mar. 2021.

\bibitem{9377456}
S.~Gu, X.~Sun, Z.~Yang, T.~Huang, W.~Xiang, and K.~Yu,
\newblock ``Energy-{Aware} coded caching strategy design with resource
  optimization for satellite-{UAV-Vehicle-Integrated} networks,''
\newblock {\em IEEE Internet of Things Journal}, vol. 9, no. 8, pp. 5799--5811,
  Mar. 2022.

\bibitem{6363993}
L.~Galluccio, G.~Morabito, and S.~Palazzo,
\newblock ``Caching in information-centric satellite networks,''
\newblock in {\em Proc. of IEEE International Conference on Communications
  (ICC)}, Ottawa,Canada, Jun. 2012, pp. 3306--3310.

\bibitem{1264069}
A.~Armon and H.~Levy,
\newblock ``Cache satellite distribution systems: modeling, analysis, and
  efficient operation,''
\newblock {\em IEEE Journal on Selected Areas in Communications}, vol. 22, no.
  2, pp. 218--228, 2004.

\bibitem{liu2020analysis}
X.~Liu,
\newblock ``Analysis in big data of satellite communication network based on
  machine learning algorithms,''
\newblock {\em Transactions on Emerging Telecommunications Technologies}, vol.
  32, no. 7, pp. e3861, Jan. 2021.

\bibitem{8353863}
A.~Gharanjik, M.~R.~B. Shankar, F.~Zimmer, and B.~Ottersten,
\newblock ``Centralized rainfall estimation using carrier to noise of satellite
  communication links,''
\newblock {\em IEEE Journal on Selected Areas in Communications}, vol. 36, no.
  5, pp. 1065--1073, May 2018.

\bibitem{8713802}
P.~V.~R. Ferreira, R.~Paffenroth, A.~M. Wyglinski, T.~M. Hackett, S.~G. Bilen,
  R.~C. Reinhart, and D.~J. Mortensen,
\newblock ``Reinforcement learning for satellite communications: From {LEO} to
  deep space operations,''
\newblock {\em IEEE Communications Magazine}, vol. 57, no. 5, pp. 70--75, May
  2019.

\bibitem{pacheco2020framework}
F.~Pacheco, E.~Exposito, and M.~Gineste,
\newblock ``A framework to classify heterogeneous internet traffic with machine
  learning and deep learning techniques for satellite communications,''
\newblock {\em Computer Networks}, vol. 173, pp. 107213, May 2020.

\bibitem{na2018distributed}
Z.~Na, Z.~Pan, X.~Liu, Z.~Deng, Z.~Gao, and Q.~Guo,
\newblock ``Distributed routing strategy based on machine learning for {LEO}
  satellite network,''
\newblock {\em Wireless Communications and Mobile Computing}, vol. 2018, Jan.
  2018.

\bibitem{8910638}
B.~Deng, C.~Jiang, H.~Yao, S.~Guo, and S.~Zhao,
\newblock ``The next generation heterogeneous satellite communication networks:
  Integration of resource management and deep reinforcement learning,''
\newblock {\em IEEE Wireless Communications}, vol. 27, no. 2, pp. 105--111,
  Nov. 2020.

\bibitem{9457160}
Y.~Hu, M.~Chen, W.~Saad, H.~V. Poor, and S.~Cui,
\newblock ``Distributed multi-agent meta learning for trajectory design in
  wireless drone networks,''
\newblock {\em IEEE Journal on Selected Areas in Communications}, vol. 39, no.
  10, pp. 3177--3192, 2021.

\bibitem{9562559}
M.~Chen, D.~Gündüz, K.~Huang, W.~Saad, M.~Bennis, A.~V. Feljan, and H.~V.
  Poor,
\newblock ``Distributed learning in wireless networks: Recent progress and
  future challenges,''
\newblock {\em IEEE Journal on Selected Areas in Communications}, vol. 39, no.
  12, pp. 3579--3605, Dec. 2021.

\bibitem{vanschoren2018meta}
J.~Vanschoren,
\newblock ``Meta-learning: A survey,''
\newblock {\em arXiv preprint arXiv:1810.03548}, 2018.

\bibitem{nichol2018first}
A.~Nichol, J.~Achiam, and J.~Schulman,
\newblock ``On first-order meta-learning algorithms,''
\newblock {\em arXiv preprint arXiv:1803.02999}, 2018.

\bibitem{chen2016caching}
M.~Chen, M.~Mozaffari, W.~Saad, C.~Yin, M.~Debbah, and C.~S. Hong,
\newblock ``Caching in the sky: {P}roactive deployment of cache-enabled
  unmanned aerial vehicles for optimized quality-of-experience,''
\newblock {\em IEEE Journal on Selected Areas on Communications (JSAC), Special
  Issue on Human-In-The-Loop Mobile Networks}, vol. 35, no. 5, pp. 1046--1061,
  May 2017.

\bibitem{tse2005fundamentals}
D.~Tse and P.~Viswanath,
\newblock {\em Fundamentals of wireless communication},
\newblock Cambridge university press, 2005.

\bibitem{puterman2014markov}
M.~Puterman,
\newblock {\em Markov decision processes: discrete stochastic dynamic
  programming},
\newblock John Wiley \& Sons, 2014.

\bibitem{watkins1992q}
C.~J. Watkins and P.~Dayan,
\newblock ``Q-learning,''
\newblock {\em Machine learning}, vol. 8, no. 3-4, pp. 279--292, 1992.

\bibitem{chen2019joint}
M.~Chen, Z.~Yang, W.~Saad, C.~Yin, H.~V. Poor, and S.~Cui,
\newblock ``A joint learning and communications framework for federated
  learning over wireless networks,''
\newblock {\em IEEE Transactions on Wireless Communications}, vol. 20, no. 1,
  pp. 269--283, Jan. 2021.

\bibitem{sutton2000policy}
R.~S Sutton, D.~A McAllester, S.~P Singh, and Y.~Mansour,
\newblock ``Policy gradient methods for reinforcement learning with function
  approximation,''
\newblock in {\em Proc. of Advances in Neural Information Processing Systems
  (NIPS)}, Denver, USA, Dec. 2000.

\bibitem{sunehag2017value}
P.~Sunehag, G.~Lever, A.~Gruslys, W.~M. Czarnecki, V.~Zambaldi, M.~Jaderberg,
  M.~Lanctot, N.~Sonnerat, J.~Z Leibo, K.~Tuyls, et~al.,
\newblock ``Value-decomposition networks for cooperative multi-agent
  learning,''
\newblock {\em arXiv preprint arXiv:1706.05296}, 2017.

\bibitem{finn2017model}
C.~Finn, P.~Abbeel, and S.~Levine,
\newblock ``Model-agnostic meta-learning for fast adaptation of deep
  networks,''
\newblock in {\em Proc. of International Conference on Machine Learning
  {(ICML)}}, Sydney, Australia, Aug. 2017.

\bibitem{erhan2010does}
D.~Erhan, A.~Courville, Y.~Bengio, and P.~Vincent,
\newblock ``Why does unsupervised pre-training help deep learning?,''
\newblock in {\em Proceedings of the thirteenth international conference on
  artificial intelligence and statistics}. JMLR Workshop and Conference
  Proceedings, 2010, pp. 201--208.

\bibitem{baird1993advantage}
L.~C. Baird~III,
\newblock ``Advantage updating,''
\newblock Tech. {R}ep., WRIGHT LAB WRIGHT-PATTERSON AFB OH, 1993.

\bibitem{sutton2018reinforcement}
R~S Sutton and A~G Barto,
\newblock {\em Reinforcement learning: An introduction},
\newblock MIT press, 2018.

\bibitem{Tracker}
``Starlink coverage tracker,'' https://starlink.sx/.

\bibitem{Update}
``Starlink daily coverage estimates,''
  https://sebsebmc.github.io/starlink-coverage/index.html.

\bibitem{foerster2017counterfactual}
J.~Foerster, G.~Farquhar, T.~Afouras, N.~Nardelli, and S.~Whiteson,
\newblock ``Counterfactual multi-agent policy gradients,''
\newblock in {\em Proceedings of the AAAI Conference on Artificial
  Intelligence}, 2018, vol.~32.

\end{thebibliography}
\end{document}